\documentclass[twocolumn]{aastex701}

\newcommand{\mathbfit}[1]{\textbf{\textit{#1}}}

\usepackage{amsmath}
\usepackage[T1]{fontenc}
\usepackage[utf8]{inputenc}
\begin{document}

\title{3D and 2D Radiation Non-ideal MHD modeling of the decay of FUor outbursts}

\author[orcid=0000-0003-2929-1502]{Yisheng Tu}
\affiliation{Department of Astronomy, University of Michigan, Ann Arbor, MI, 48109}
\email{yitu@umich.edu}

\author[orcid=0000-0003-1430-8519]{Lee Hartmann} 
\affiliation{Department of Astronomy, University of Michigan, Ann Arbor, MI, 48109}
\email{lhartm@umich.edu}

\author[orcid=0000-0003-3616-6822]{Zhaohuan Zhu}
\affiliation{Department of Physics and Astronomy, University of Nevada, Las Vegas, 89154}
\affiliation{Nevada Center for Astrophysics, University of Nevada, Las Vegas, 4505 S. Maryland Pkwy, Las Vegas, NV, 89154}
\email{zhaohuan.zhu@unlv.edu}

\author[orcid=
0000-0002-8558-1427]{Conel M. O'D. Alexander}
\affiliation{
Earth and Planets Laboratory, Carnegie Institution for Science, Washington DC, 20015}
\email{calexander@carnegiescience.edu}

\author[orcid=
0000-0002-0093-065X]{Fred J. Ciesla}
\affiliation{Department of the Geophysical Sciences,
University of Chicago,
Chicago, IL, 60637}
\email{fciesla@uchicago.edu}

\author[orcid=0009-0006-7444-6509]{Eric Rapp III}
\affiliation{Department of Astronomy, University of Michigan, Ann Arbor, MI, 48109}
\email{ericrapp@umich.edu}



\begin{abstract}
FU Orionis (FUor) outbursts are high-accretion events in protoplanetary disks thought to be driven by the magnetorotational instability (MRI). 
Although these outbursts span a wide range of observed durations and can thermally process and potentially transport disk material, 
the physical processes governing their decay remain poorly understood. We present a three-dimensional radiative non-ideal magnetohydrodynamic (MHD) simulation of the decay of an MRI-driven FUor outburst, complemented by a suite of two-dimensional models that explore its dependence on disk properties. During the decay, rings and gaps form naturally within the inner disk. Magnetic flux preferentially concentrates in the gaps, while the rings remain comparatively weakly magnetized. We show that MRI-active rings act as ``magnetic dams,'' confining magnetic flux within adjacent gaps. As the rings cool below the MRI activation temperature, non-ideal MHD effects decouple the gas from the magnetic field, allowing the trapped flux to diffuse outward. This outward diffusion can sometimes transiently reactivate MRI in neighboring rings, producing localized reheating events that we term ``Reactivation.'' By comparing otherwise identical two- and three-dimensional simulations, we find that MRI heating is systematically less efficient in two dimensions, leading to outburst durations shorter by a factor of $\sim3$--4. Our model suite further shows that the decay timescale depends primarily on the disk surface density and distance from the central star, with only a weak dependence on the magnetic field strength. We further derive an empirical relation linking the outburst duration to these quantities.
\end{abstract}

\keywords{\uat{Stellar accretion disks}{1579} --- \uat{FU Orionis stars}{553} --- \uat{Star formation }{1569} --- \uat{Magnetohyrodynamical simulations}{1857} --- \uat{Radiative magnetohydrodynamics}{2009}}


\section{Introduction} 
FU Orionis-type objects (FUors) represent one of the most dramatic manifestations of episodic accretion during the early stages of star formation. These systems undergo sudden luminosity increases of several magnitudes, corresponding to accretion rates rising from typical T Tauri values of $\sim10^{-8}$--$10^{-7}~M_\odot~\mathrm{yr}^{-1}$ to as high as $\sim10^{-4}~M_\odot~\mathrm{yr}^{-1}$, with elevated accretion persisting for decades to centuries \citep{Hartmann1996,Audard2014,Fischer2023}. Over the past decades, an increasing number of FUor and FUor-like systems have been identified across a broad range of protostellar evolutionary stages \citep[e.g.,][]{Hillenbrand2018, Hillenbrand2022, Hillenbrand2023, Contreras2023, Contreras2025, Giannini2024, Morris2025}. Statistical arguments further suggest that most low-mass stars likely undergo multiple such outbursts during their assembly history, implying that episodic accretion may be a nearly universal phase of low-mass star and planet formation \citep[][]{Scholz2013, Audard2014, Hartmann2025, Carvalho2026}.

The importance of FUor outbursts extends far beyond the transient brightening event itself. Episodic accretion has long been proposed as a solution to the ``luminosity problem'' in protostellar evolution, in which observed protostars appear systematically underluminous compared with predictions from steady-accretion models \citep[][]{Kenyon1990, Evans2009, Offner2011, Dunham2012}. Outbursts also substantially reshape the thermal and chemical structure of protostellar disks and envelopes, potentially altering dust evolution, volatile chemistry, and the initial conditions for planet formation \citep{cieza2016,
Molyarova2018, Lee2019}. In addition, FUor events are frequently associated with enhanced or altered winds and jets, indicating a close interplay among accretion, magnetic fields, and outflow launching \citep[e.g.,][]{Konigl2011, Principe2018, Cruz-Saenz2023}. In the context of Solar System formation, FUor-like outbursts may also provide a natural explanation for the presence of thermally processed material in comets and crystalline material in outer disks that are believed to have formed far from the Sun \citep{Nuth2006, Ciesla2009, Abraham2009}.

Despite decades of study, the physical mechanism responsible for triggering FUor outbursts remains uncertain. Many possibilities have
have been proposed, such as thermal instability \citep[TI;][]{Bell1994}, tidal perturbations induced by stellar companions or flybys \citep{Borchert2022}, along with
others (\citealt{Vorobyov2020, Kadam2020, Demidova2023}; see discussion in 
\citealt{Vorobyov2021}).
One of the most common suggestions
is heating by gravitational instability (GI) that leads to activating the outburst via the onset of the magnetorotational instability (MRI) \citep{armitage01,
Zhu2009, Zhu2010a, Zhu2010b, Bae2013, Dong2016,Bourdarot2023},
although it may be difficult for GI to excite short bursts \citep{Cleaver2023}.
In any case, it is likely that the MRI is driving accretion during outbursts,
as disk temperatures are high enough to maintain enough ionization for magnetic activation, and magnetic fields must be present to explain the cold high-velocity winds of FUors \citep{Hartmann1996}.
Understanding the decay of the outburst, and investigating the possible post-outburst effects on disk materials and structure, thus requires an exploration of the way how MRI shuts off.

In many existing models, the decay phase is treated phenomenologically or uses simplified prescriptions for the effects of magnetic field (typically as viscosity), non-ideal MHD, and gas thermal evolution \citep[][]{Martin2012, Martin2014, Lin2014, Kadam2020, Vorobyov2020}. In reality, the transition from the high-accretion state back to quiescence likely involves a complex interplay among disk thermodynamics, magnetic fields, radiation, and non-ideal magnetohydrodynamic (MHD) effects. In particular, the efficiency of magnetic coupling depends sensitively on the thermal structure of the disk, while the thermal evolution itself is strongly influenced by MRI-driven heating and radiative cooling \citep[as discussed in e.g.,][]{Zhu2020, Lohnert2022}. Capturing these coupled processes self-consistently is therefore essential for understanding the termination of FUor outbursts.


In this work, we present a 3D radiation non-ideal MHD simulation for
the decay of the MRI in a rapidly-accreting protoplanetary disk model, and analyze its properties in detail.
While such a 3D calculation is indispensable for capturing the inherently multidimensional dynamics of the system, these are computationally very expensive, making
extensive parameter studies prohibitive. We therefore use long-duration 2D simulations to explore and generalize the physical picture,
examining the 
diversity of outburst decay behavior across different system parameters,
starting from a detailed comparison of a 2D version of the 3D model.

This paper is organized as follows. In Section~\ref{sec:method}, we describe the simulation setup and model parameters. In Section~\ref{sec:3D_results}, we present a detailed analysis of the three-dimensional simulation, followed by a parameter study and extrapolation using two-dimensional models in Section~\ref{sec:2D}. We discuss the implications and limitations of our results in Section~\ref{sec:discussion}, and conclude in Section~\ref{sec:conclusion}.

\section{Method}
\label{sec:method}
The evolution of the inner disk is governed by the coupled non-ideal MHD equations 
\begin{equation}
    \frac{\partial \rho}{\partial t}+\nabla\cdot(\rho\mathbfit{v}) = 0,
    \label{equ:rmhd:mass}
\end{equation}
\begin{equation}
    \frac{\partial\rho\mathbfit{v}}{\partial t} + \nabla\cdot\Big[\rho\mathbfit{v}\mathbfit{v} + \big(P+\frac{B^2}{8\pi}\big)\mathbfit{I} - \frac{\mathbfit{B}\mathbfit{B}}{4\pi}\Big] = -\mathbfit{S}_r(\mathbfit{P}) - \rho\mathbfit{g},
    \label{equ:rmhd:momentum}
\end{equation}
\begin{equation}
    \frac{\partial E}{\partial t} + \nabla\cdot[(E+P)\mathbfit{v} - \mathbfit{B}(\mathbfit{B}\cdot\mathbfit{v})+\mathbfit{S}_\mathrm{ni}] = - \rho\mathbfit{g}\cdot\mathbfit{v} - cS_r(E),
    \label{equ:rmhd:energy}
\end{equation}
\begin{equation}
    \frac{\partial \mathbfit{B}}{\partial t} = \nabla\times(\mathbfit{v}\times\mathbfit{B}) - \frac{4\pi}{c}\nabla\times[\eta_O\mathbfit{J} + \eta_A\mathbfit{J}_\perp],
    \label{equ:rmhd:magnetic}
\end{equation}
\begin{equation}
    \frac{dI}{dt} +c\mathbfit{n}\cdot\nabla I=cS_I.
    \label{equ:rmhd:rad_energy}
\end{equation}
Here, $\rho$, $P$, $\mathbfit{v}$, and $\mathbfit{B}$ represent the gas density, thermal pressure, velocity, and magnetic field, respectively. $\mathbfit{I}$ is the identity matrix. The total energy density is given by $E=E_g+\rho \mathbfit{v}^2/2+\mathbfit{B}^2/8\pi$, where $E_g$ is the internal energy density and $\mathbfit{S}_\mathrm{ni}$ is the source term due to non-ideal MHD effects. The gravitational acceleration $g$ is the acceleration due to the central star. Because the gravitational potential in the inner disk is dominated by the central star and the inner disk is expected to be gravitationally stable (Toomre-$Q\gg1$ for $R<1~\mathrm{au}$ in all our models), we neglect disk self-gravity to reduce the computational cost. The magnetic current density is defined as $\mathbfit{J}=(c/4\pi)\nabla\times\mathbfit{B}$, and its component perpendicular to the magnetic field is $\mathbfit{J}_\perp=[(\mathbfit{J}\times\mathbfit{B})\times\mathbfit{B}]/B^2$. 
The Ohmic and Ambipolar diffusivities are
$\eta_O$ and $\eta_A$; we neglect the Hall effect as a first step.
$I$ is the frequency-integrated specific intensity and is used to compute the radiative source terms $S_I$, $cS_r(E)$, and $\mathbfit{S}_r(\mathbfit{P})$ following \citet{Jiang2014, Jiang2021}. For computational feasibility, we multiply the speed of light by $3\times10^{-3}$, which substantially reduces the computational cost while having minimal impact on the simulation outcome \citep[][]{Zhu2020}.

The system of equations is solved using \texttt{Athena++} in spherical-polar coordinates \citep{Stone2020, Jiang2014}. To model the outbursting inner disk while maintaining a reasonable simulation time step and allowing the full temporal evolution of the outburst decay to be captured, the simulation domain of the reference 3D model spans $0.1<r<10~\mathrm{au}$, $0<\theta<\pi$, and $0<\phi<\pi/2$, with a base resolution of $(N_r\times N_\theta\times N_\phi)=(80\times64\times32)$. The grid is logarithmically spaced in $r$ and linearly spaced in both $\theta$ and $\phi$, yielding cells that are approximately cubic throughout the domain.

To adequately resolve the inner disk while maintaining a reasonable computational cost, we employ three levels of mesh refinement, each covering the full azimuthal extent. The highest refinement level spans $0.1<r<0.5~\mathrm{au}$, where MRI activity is strongest. Initially, this refinement provides approximately eight cells per scale height at $R=0.4~\mathrm{au}$ and five cells per scale height at the inner radial boundary. A resolution study is presented in Appendix~\ref{sec:convergence}, demonstrating that our main conclusions are not sensitive to numerical resolution. 

In this study, we focus on the decay of an outburst, so we initialize the simulation assuming an outburst has already commenced, with a relatively large initial active zone ($\approx0.6~\mathrm{au}$ after initial relaxation; see section~\ref{sec:3D_results}). The initial density and temperature profiles are prescribed as \citep[following][]{Zhu2020}
\begin{equation}
    \rho(R)=\rho_0\left(\frac{R}{R_0}\right)^p,
    \qquad
    T(R)=T_0\left(\frac{R}{R_0}\right)^q,
    \label{equ:init_rho_T}
\end{equation}
where $R_0=0.1~\mathrm{au}$, $\rho_0=1.9 \times 10^{-8}~\mathrm{g~cm^{-3}}$, $p=-1.7$, $T_0=3000~\mathrm{K}$, and $q=-0.48$. The initial temperature depends only on the cylindrical radius and is assumed to be uniform in both the vertical and azimuthal directions. The density throughout the rest of the domain is calculated by assuming vertical hydrostatic equilibrium. These parameters yield a total disk mass of $0.017~M_\odot$ within the computational domain.
The central star has a mass of $1 M_{\odot}$, a stellar surface temperature $T_\star=5666~\mathrm{K}$ and a stellar radius $r_\star=3R_\odot$, resulting in a luminosity of $8.4 ~L_{\odot}$. To ensure numerical stability, we apply a density floor following
\begin{equation}
    \rho_\mathrm{floor}(r)= 6\times10^{-18}~\mathrm{g~cm^{-3}}\times \max\Big[1, \Big(\frac{r}{r_\mathrm{floor}}\Big)^{-2}\Big],
\end{equation}
where $r_\mathrm{floor}=0.2~\mathrm{au}$.

The initial velocity field corresponds to a Keplerian disk corrected for radial pressure support. The initial magnetic field is purely vertical, with a midplane plasma-$\beta=1000$ at all cylindrical radii. We include Ohmic dissipation and ambipolar diffusion using the tabulated coefficients of \citet{Desch2015}, while neglecting the Hall effect as a first step. The tabulated coefficients require four inputs: gas density, temperature, plasma-$\beta=P_{\rm gas}/P_{\rm mag}$, and the non-thermal ionization rate $\zeta$. The first three are obtained directly from the simulation, while $\zeta=\zeta_d+\zeta_i$ includes contributions from radioactive decay and cosmic rays, respectively. Following \citet{Turner2009}, we adopt $\zeta_d=3.7\times10^{-19}~\mathrm{s^{-1}}$ and calculate the cosmic-ray contribution using the attenuation prescription of \citet{Umebayashi2009}, $\zeta_i = 10^{-17}\exp(-\Sigma/\Sigma_0)\ \mathrm{s^{-1}}$, where $\Sigma$ is the vertical column density above the cell and $\Sigma_0=96~\mathrm{g~cm^{-2}}$. For computational efficiency, particularly in dense, optically thick regions, we approximate $\Sigma\approx\rho H$, where $H$ and $\rho$ are the local disk scale height and gas density, respectively. As a first step, we neglect X-ray ionization in the main model suite because X-rays can be heavily shielded by the gas interior to $0.1~\mathrm{au}$ during an outburst, and it is computationally intensive to treat self-consistently. We further show in Appendix~\ref{sec:X-ray} that including X-ray ionization has only a minimal impact on the simulation results.

The opacity table is adopted from \citet{Zhu2021}, assuming a dust-to-gas mass ratio of 0.01 \citep[see also][]{Birnstiel2018}. At the inner radial boundary, we employ a modified outflow condition that permits mass to leave the computational domain (accretion onto the star) while minimizing inflow. Standard outflow boundary conditions are applied at the outer radial boundary. Polar boundary conditions are used at both $\theta$ boundaries, and periodic boundary conditions are adopted in the azimuthal direction.
\subsection{Suite of 2D models}
The 3D models are very computationally expensive -- self-consistent radiation non-ideal MHD is time-consuming. 
We therefore use 2D models to explore parameter space with a more reasonable computational time and cost, particularly for the parameter space with a very long decay time. These 2D simulations adopt the same $(r,\theta)$ grid structure as the 3D model, and share the same setup except for the density and/or magnetic parameter specified below.  

As a first step in exploring parameter space, we present a suite of four 2D models, listed below
\begin{enumerate}
    \item D1B1(2D) model: identical parameters as in the 3D model. This serves as a baseline for interpreting the results of the 2D models (section~\ref{sec:compare_2D_3D});
    \item D3B1 model: three times higher density than the reference model ($\rho_0$ increased by a factor of 3), keeping the same midplane plasma-$\beta=1000$;
    \item D3B3 model: three times higher density than the reference model, and midplane plasma-$\beta=3000$;
    \item D9B3 model: nine times higher density than in the reference model, and midplane plasma-$\beta=3000$.
\end{enumerate}
These 2D models are discussed in detail in section~\ref{sec:2D}. Four stress test models with a different density slopes are discussed in Appendix~\ref{app:stress}.

\begin{figure*}
    \centering
    \includegraphics[width=0.8\linewidth]{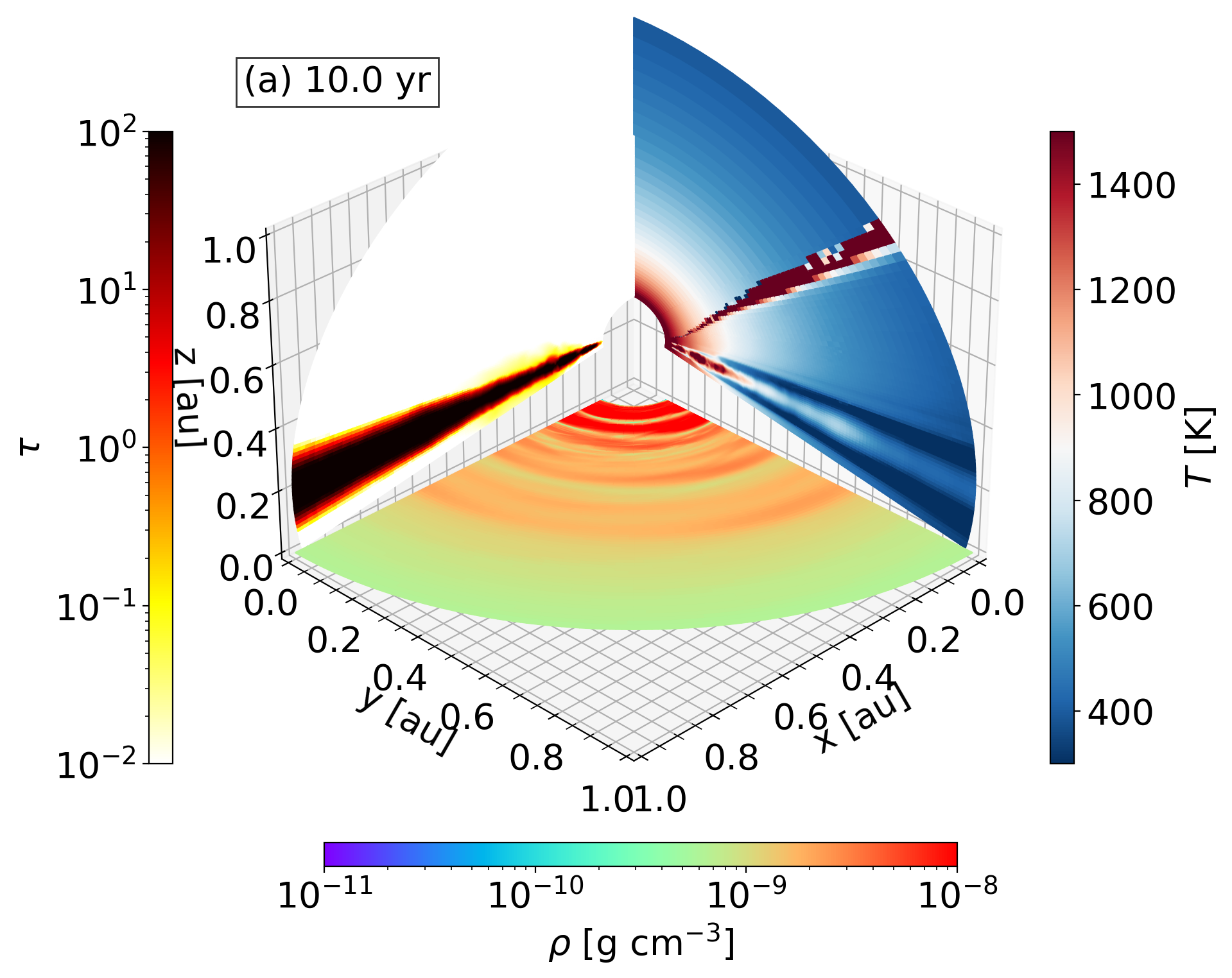}
    \caption{Three-dimensional rendering of the simulation domain. The left panel shows the vertical Rosseland mean optical depth $\tau_R$, illustrating the optically thin envelope surrounding the optically thick disk. The right panel shows a vertical temperature slice at $\phi=0$. The high-temperature wedge visible in the envelope is a numerical artifact and has a negligible impact on the simulation evolution. The bottom panel shows the midplane gas density distribution, highlighting the formation of rings and gaps.}
    \label{fig:f3D}
\end{figure*}

\section{3D Results}
\label{sec:3D_results}

We begin with an overview of the main features of the 3D model.
Figure~\ref{fig:f3D} presents representative slices of the simulation domain in a snapshot at an elapsed time of 10 yr. The lower panel shows the midplane density, where substructures have already emerged. 
The left panel shows the vertically integrated optical depth measured from the simulation boundary. The sharp contrast between the disk and surrounding atmosphere is evident: the disk remains optically thick ($\tau \gg 1$) over most of its extent, while the surrounding atmosphere is predominantly optically thin ($\tau \ll 1$). This structure highlights the distinct radiative regimes of the disk and envelope that characterize the simulated evolution.
The right panel shows the temperature structure, highlighting a heated inner disk extending to $\lesssim0.5~\mathrm{au}$ at the time shown, surrounded by a cooler outer disk. We note that the high-temperature wedge visible in the envelope 
is mainly due to the density floor applied in the envelope, and has a negligible effect on the overall evolution of the simulation. As the simulation evolves, this heated inner region gradually diminishes, primarily due to radiative cooling (see Section~\ref{sec:heating_and_cooling}). Minor numerical noise associated with mesh refinement and sharp density gradients in the envelope is present but does not affect the main results.

\begin{figure*}
    \centering
    \includegraphics[width=\linewidth]{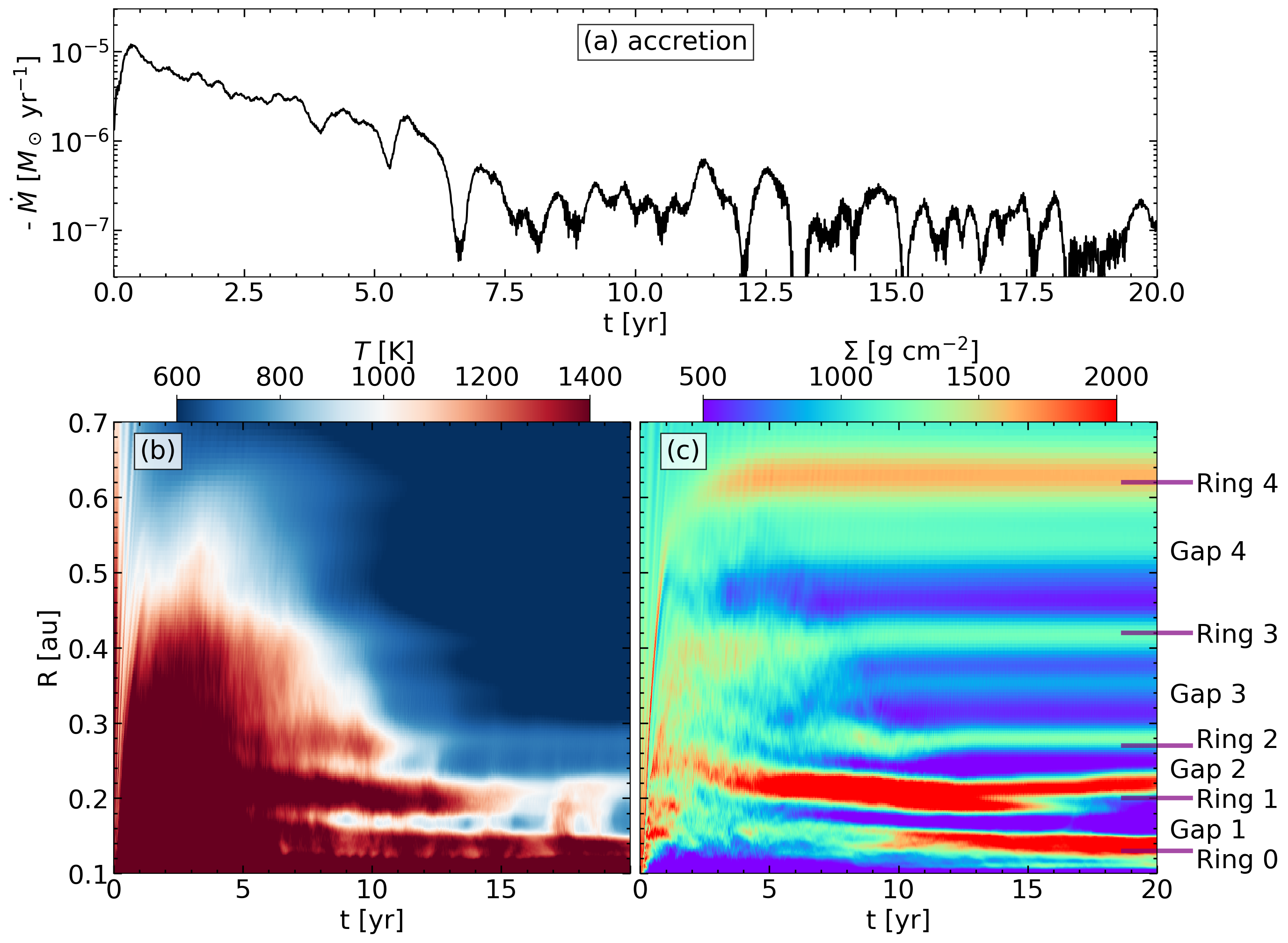}
    \caption{\textbf{Panel (a):} accretion rate in the 3D simulation, defined as the mass flux through the $R=0.105~\mathrm{au}$ sphere. \textbf{Panel (b) and (c)} show the time-space diagram of vertically-averaged temperature and column density, respectively. The labelings of the rings and gaps are shown to the right of panel (c), and the naming is chosen so that Ring $n$ is immediately exterior to Gap $n$.}
    \label{fig:spacetime_and_accretion}
\end{figure*}

To further characterize the global evolution, Figure~\ref{fig:spacetime_and_accretion}(a) shows the mass accretion rate, defined as the mass flux through $r=0.105~\mathrm{au}$. Panel (b) presents the time–space diagram of the azimuthally and vertically density-weighted averaged temperature within the disk, which is defined using the radial radiation flux criteria ($F_r<0.1F_{r,\star}$; see section~\ref{sec:heating_and_cooling}), and panel (c) shows the azimuthally averaged column density. Following initialization, the disk undergoes a rapid cooling phase, after which the inner region ($R\sim0.1$–$0.7~\mathrm{au}$) reheats from the inside out over $\sim1$~yr, while the outer disk remains cool. This behavior reflects the system adjusting to the initial conditions. Subsequently, the temperature between $R\sim0.3~\mathrm{au}$ and $0.7~\mathrm{au}$ gradually declines, falling below $\sim800~\mathrm{K}$ by $t\sim12$~yr. In contrast, the inner region ($R\lesssim0.3~\mathrm{au}$) develops three ring-like enhancements in column density (Figure~\ref{fig:spacetime_and_accretion}[c]), and exhibits some interesting dynamics as
discussed in \S 
~\ref{sec:magfield_evol_ring_decay} and \ref{sec:terminal_reactivation}.




\subsection{Rings and gaps}
\label{sec:rings_and_gaps}
\begin{figure*}
    \centering
    \includegraphics[width=\linewidth]{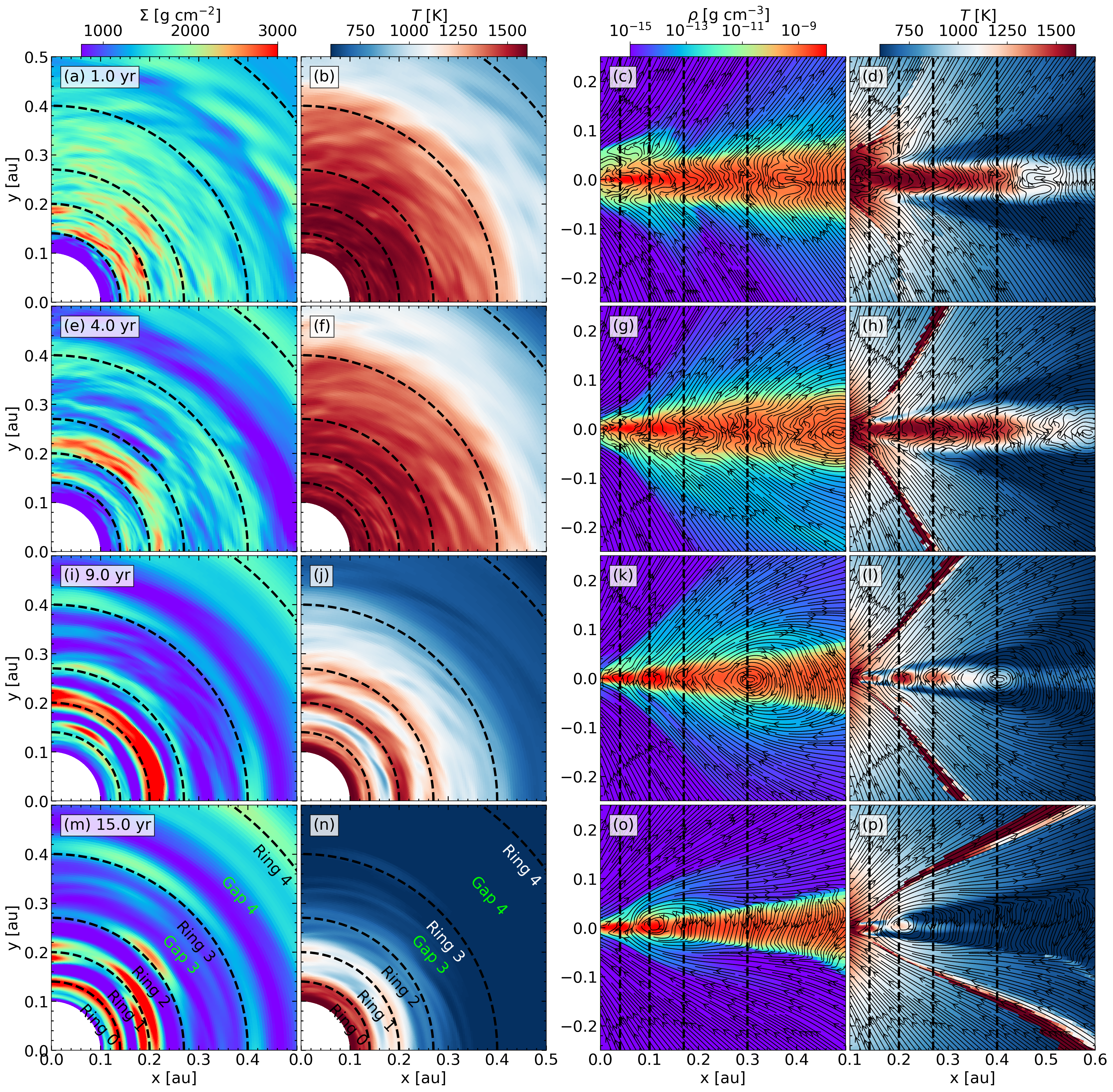}\caption{Evolution of density and temperature in the simulation. The left two columns (\textbf{Panels a, b, e, f, i, j, m, and n}) show the evolution of column density and vertically-averaged temperature; The right two columns (\textbf{Panels c, d, g, h, k, l, o, p}), show the azimuthal-averaged density and temperature, overplotted by azimuthal-averaged magnetic field lines. The four vertical dashed lines show the locations of the innermost four rings, as labeled in figure~\ref{fig:spacetime_and_accretion} and panels (m) and (n) here. The diagonal high-temperature wings in the rightmost column are numerical artifacts that do not affect the results of the simulation.}
    \label{fig:ring_gap}
\end{figure*}

For ease of reference in the following discussion, we adopt the ring nomenclature indicated in Figure~\ref{fig:spacetime_and_accretion}[c] (see also Figure~\ref{fig:ring_gap}[m] and [n]). The innermost ring, located at $\sim 0.12~\mathrm{au}$, is labeled ``Ring 0,'' followed by ``Ring 1'' at $0.2~\mathrm{au}$, and ``Ring 2,'' ``Ring 3,'' and ``Ring 4'' at progressively larger radii. The gaps are labeled accordingly: ``Gap 1'' lies between Rings 0 and 1, ``Gap 2'' between Rings 1 and 2, and so on. By construction, Gap $n$ is always immediately interior to Ring $n$. 


To illustrate the formation and evolution of these structures, Figure~\ref{fig:ring_gap} shows the density and temperature distributions in the disk. Early in the simulation, after the initial adjustment phase, weak substructures—characterized by slight over- and under-densities and minor temperature fluctuations—begin to emerge (Figure~\ref{fig:ring_gap}[a–d], at $t=1~\mathrm{yr}$). These perturbations subsequently grow, leading to increasingly pronounced surface density contrasts (Figure~\ref{fig:ring_gap}[e–h]). As the system evolves further, gas becomes concentrated into more coherent and well-defined rings (Figure~\ref{fig:ring_gap}[i–l]). By the end of the simulation, the disk exhibits approximately five distinct rings in column density (Figure~\ref{fig:ring_gap}[m–p]).

The density enhancements associated with the rings and gaps persist even after these regions have cooled and become decoupled from the magnetic field. For example, by $t=4~\mathrm{yr}$, Ring 4 has already cooled, yet it remains clearly identifiable in $\Sigma$ throughout the simulation. Similarly, Ring 3 is already cooled by $t=9~\mathrm{yr}$ (Figure~\ref{fig:ring_gap}[i]), and Ring 2 by $t=15~\mathrm{yr}$, but in both cases the density structures persist. Further implications of these long-lived rings are discussed in Section~\ref{sec:magfield_evol_ring_decay} and \ref{sec:long_live_aftermath}.

Given the central role of magnetic fields in governing the system’s evolution, the right panels of Figure~\ref{fig:ring_gap} present the azimuthally averaged density and temperature profiles. Although the simulation is initialized with a purely vertical magnetic field, the field lines are rapidly bent and advected across the disk. At later times (particularly after $\sim 9~\mathrm{yr}$), the field near the disk becomes strongly bent backward, forming closed meridional loops whose outer footpoints typically reside in the cooler, magnetically dead zone. This field configuration is physically expected: part of the field remains anchored in the well-coupled inner disk (where temperatures exceed the MRI activation threshold), while the outward bending transports magnetic flux into cooler, less ionized regions. In these outer regions, weaker coupling allows the field lines to be pushed into the disk and form closed loops. This geometry has important implications for the magnetic flux budget (\S ~\ref{sec:magfield_evol_ring_decay}).

We note that ring--gap formation is a common outcome of MHD disk simulations and that our results are consistent with this general behavior in previous studies. Both local shearing-box models \citep[e.g.,][]{johansen2009,Riols2018,Riols2019,Riols2020} and global simulations \citep[][ see also \citealt{Lesur2023}]{Flock2015, Suriano2018} have found the spontaneous development of rings, often referred to as ``zonal flows''. Non-ideal MHD effects are not required for ring formation, although they can facilitate the process by modifying the coupling between the gas and magnetic field \citep[][]{Steinacker2002, Bai2014}. In our simulations, the inner disk is initially hot and therefore well coupled to the magnetic field. As the disk cools and rings begin to emerge, however, non-ideal MHD effects become increasingly important and likely assist the development of the ring-gap structures, consistent with previous studies. More importantly, non-ideal MHD effects play a second role unique to the outburst decay phase: they regulate the long-term evolution of the magnetic flux by allowing magnetic fields concentrated in gaps to leave the inner disk once the rings cool below the MRI activation temperature. This process governs the subsequent evolution of the rings and ultimately facilitates the decay of the outburst, as discussed in detail in Sections~\ref{sec:magfield_evol_ring_decay} and \ref{sec:terminal_reactivation}.
 

\subsection{Disk Thermodynamics}
\label{sec:heating_and_cooling}
The thermodynamic evolution of the disk is governed by the competition between magnetic heating and radiative cooling. We first quantify the cooling rate associated with radiation. This is not straightforward because the radiation field contains contributions from both intrinsic disk cooling and stellar irradiation. Although the radiative flux is directly available from the simulation, care must be taken when using it to estimate the cooling rate. The dominant energy loss from the disk occurs through the polar-direction radiative flux, $F_\theta$, since the disk surface is approximately normal to the $\hat{\theta}$ direction. However, evaluating $F_\theta$ too high above the disk places the measurement in the optically thin envelope, where the radiative flux can be dominated by scattered stellar irradiation and therefore overestimates the intrinsic cooling of the disk. Conversely, evaluating it too deeply inside the optically thick disk would underestimate the total escaping radiative flux.

To define an appropriate emitting surface, we instead use the transition between the optically thin envelope and the optically thick disk. In the envelope, the radial radiative flux is approximately equal to the stellar irradiation flux,
\begin{equation}
    F_r \approx F_{r,\star}
    = \sigma T_\star^4\left(\frac{r}{r_\star}\right)^{-2},
\end{equation}
whereas deep inside the disk the radial flux becomes negligible ($F_r\approx0$). We therefore define the effective emitting surface as the location where
$F_r = 0.1~F_{r,\star},$
which approximately marks the transition from stellar-irradiation-dominated radiation to disk-dominated radiation. We note that this definition yields a surface similar to that defined by the radial optical-depth criterion, $\tau_r \sim 1$, and that either definition produces similar results. The radiative cooling rate is then estimated as
\begin{equation}
    \frac{d\bar{\mathcal{E}}}{dt}\Big|_\mathrm{rad}
    = \sum_\mathrm{upper,~lower} F_\theta\left(r|F_r=0.1F_{r,\star}\right),
    \label{equ:dedt_rad}
\end{equation}
where the sum is over the disk's upper and lower surfaces to account for radiative flux loss through both sides of the disk.

This definition provides a practical estimate of the disk cooling rate while minimizing contamination from direct stellar irradiation. Although the choice of the threshold ($0.1~F_{r,\star}$) is somewhat empirical, we find that the inferred cooling rate is insensitive to modest changes in this value.

While magnetic heating is naturally included in the energy equation (equation~\ref{equ:rmhd:energy}), its overall effect is difficult to estimate directly. We therefore adopt the standard $\alpha$ prescription to capture its overall contribution and facilitate comparison with previous models (section~\ref{sec:discussion}). 
The parameter $\alpha$ represents the effective viscosity induced by Maxwell stresses\footnote{Although the Reynolds stress can in principle also contribute to the heating, we have verified that its contribution is minimal compared to the Maxwell stress.}, and is defined as
\begin{equation}
    \alpha = \frac{2}{3}\frac{\langle T_m \rangle}{\langle P_r \rangle},
    \label{equ:alphas}
\end{equation}
where
\begin{equation}
    \langle T_m \rangle = \frac{1}{2h}\int \frac{B_r B_\phi}{4\pi} dz,
    \label{equ:Tf}
\end{equation}
and
\begin{equation}
    \langle P_r \rangle = \frac{1}{2h}\int P dz.
    \label{equ:Pm}
\end{equation}
Here, $h$ is the disk half-thickness measured from the midplane, $B_r$ and $B_\phi$ are the radial and azimuthal components of the magnetic field, and $P$ is the gas thermal pressure. These gas and magnetic quantities are computed as vertically averaged values over the disk.

The corresponding heating rate is, therefore, 
\begin{equation}
    \frac{d\bar{\mathcal{E}}}{dt}\Big|_\mathrm{MRI}
    = \frac{9}{4}\alpha c_s^2 \Sigma \Omega,
    \label{equ:dedt_MRI}
\end{equation}
where $c_s$, $\Sigma$, and $\Omega$ are the gas sound speed, column density, and angular velocity, respectively. 

\begin{figure*}
    \centering
    \includegraphics[width=\linewidth]{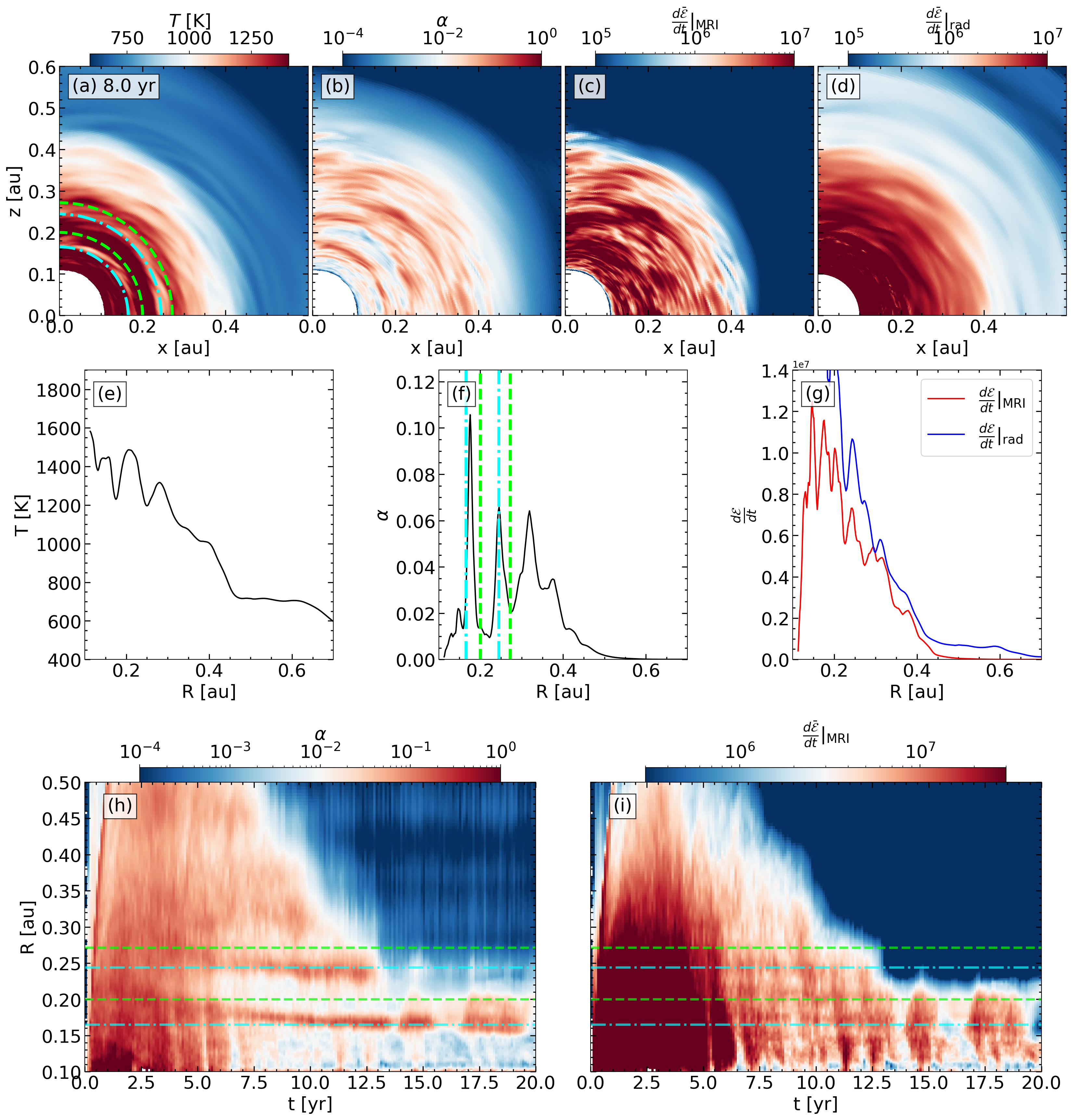}
    \caption{\textbf{Panels (a--d)} show the vertically averaged disk temperature, effective viscous $\alpha$ (equation~\ref{equ:alphas}), and the energy generation and loss rates (equations~\ref{equ:dedt_MRI} and \ref{equ:dedt_rad}), respectively. \textbf{Panels (e--g)} show the vertically and azimuthally averaged temperature, $\alpha$, and energy generation/loss rates, respectively, quantifying the thermal structure and energetics of the disk. The vertical dashed lines in panel f correspond to those shown in panel a. Green dashed lines indicate the locations of the rings, while cyan dash-dotted lines mark the locations of the gaps. \textbf{Panels (h) and (i)} show the time--space diagrams of the effective $\alpha$ and MRI energy generation rate, respectively.}
    \label{fig:heating_cooling_and_spacetime}
\end{figure*}

Figures~\ref{fig:heating_cooling_and_spacetime}(a–d) show the face-on distributions of the disk-averaged temperature, effective viscosity $\alpha$, MRI heating rate, and radiative cooling rate at a representative time $t = 8.0~\mathrm{yr}$. Since the system remains approximately axisymmetric throughout its evolution, we further present azimuthally averaged quantities in Figures~\ref{fig:heating_cooling_and_spacetime}(e–g), including the temperature, $\alpha$, and heating/cooling rates. At this time, $\alpha$ ranges from $\sim 0.01$ to $0.1$ within the MRI-active inner disk ($R \lesssim 0.5~\mathrm{au}$), and drops to negligible values in the outer, magnetically dead region ($R \gtrsim 0.5~\mathrm{au}$). As a result, the MRI heating rate is comparable to the radiative cooling rate only at the outer regions of the active zone ($\sim0.25~\mathrm{au}\lesssim r\lesssim~0.4~\mathrm{au}$). We note the net heating rate is higher than the MRI-only heating rate shown in Figure~\ref{fig:heating_cooling_and_spacetime}(c) and (g) at small radii $r\lesssim0.2~\mathrm{au}$ due to stellar irradiation heating in that region.

As shown in Figure~\ref{fig:heating_cooling_and_spacetime}(f), $\alpha$ is systematically larger within the gaps
between rings (corresponding to lower-temperature regions; cyan line in Figure~\ref{fig:heating_cooling_and_spacetime}[a]) and smaller within the rings (higher-temperature regions; green line in Figure~\ref{fig:heating_cooling_and_spacetime}[a]).
This behavior persists throughout the simulation, as shown in the time–space diagrams of $\alpha$, as well as the MRI-heating rate in Figures~\ref{fig:heating_cooling_and_spacetime}(h) and (i). Two prominent rings form: one located at $R \approx 0.28~\mathrm{au}$ (which fades around $t \approx 12~\mathrm{yr}$; Ring 2), and another at $R \approx 0.2~\mathrm{au}$ (Ring 1). These ring locations are indicated by green dashed lines in Figures~\ref{fig:heating_cooling_and_spacetime}(h) and (i), while the corresponding gaps are marked by cyan dash-dotted lines. It is evident that $\alpha$ is consistently higher in the gaps than in the rings, often by nearly an order of magnitude. However, this enhancement in $\alpha$ does not translate into a correspondingly higher MRI heating rate (Figure~\ref{fig:heating_cooling_and_spacetime}[i]). 
This is because although $\alpha$ is larger in the gaps, the surface density $\Sigma$ is higher in the rings (see Equation~\ref{equ:dedt_MRI}), resulting in comparable heating rates between rings and gaps.

An additional noteworthy feature is the time variation of the outer ring (Ring 2, at $R \approx 0.28~\mathrm{au}$) and its associated gap (Gap 2, at $R \approx 0.25~\mathrm{au}$). At earlier times, when the ring remains MRI-active, $\alpha$ is elevated in the gap. However, when the ring cools and becomes MRI-inactive at $t \approx 13~\mathrm{yr}$, the heating in the gap also diminishes simultaneously. This behavior reflects the shutdown of MRI activity within the ring as a
result of the evolution in magnetic flux, as discussed next.

\subsection{Magnetic field evolution and ring decay}
\label{sec:magfield_evol_ring_decay}
Because the 
amount of vertical magnetic flux $\Phi$ threading the disk determines the available field lines for generating $B_\phi$ and $B_r$, and
thus the amount of heating, it is instructive to examine its evolution.
Care must be taken when evaluating the flux through the disk midplane because part of the flux threads the inner radial boundary. We therefore define the midplane flux in the simulation as
\begin{equation}
    \Phi_\mathrm{mid}(R)=\oint_\mathrm{ib} \mathbfit{B}_z\cdot d\mathbfit{A} + \int_{R'=R_\mathrm{in}}^R \mathbfit{B}_z \cdot d\mathbfit{A},
    \label{equ:mag_flux}
\end{equation}
where the first term represents the vertical magnetic flux threading the inner boundary (``ib'' stands for ``inner boundary''), and the second term is the usual magnetic flux integrated over the disk midplane. This definition captures the net magnetic flux contained within a cylindrical radius and therefore tracks the gain and loss of magnetic flux in the disk.

\begin{figure}
    \centering
    \includegraphics[width=\linewidth]{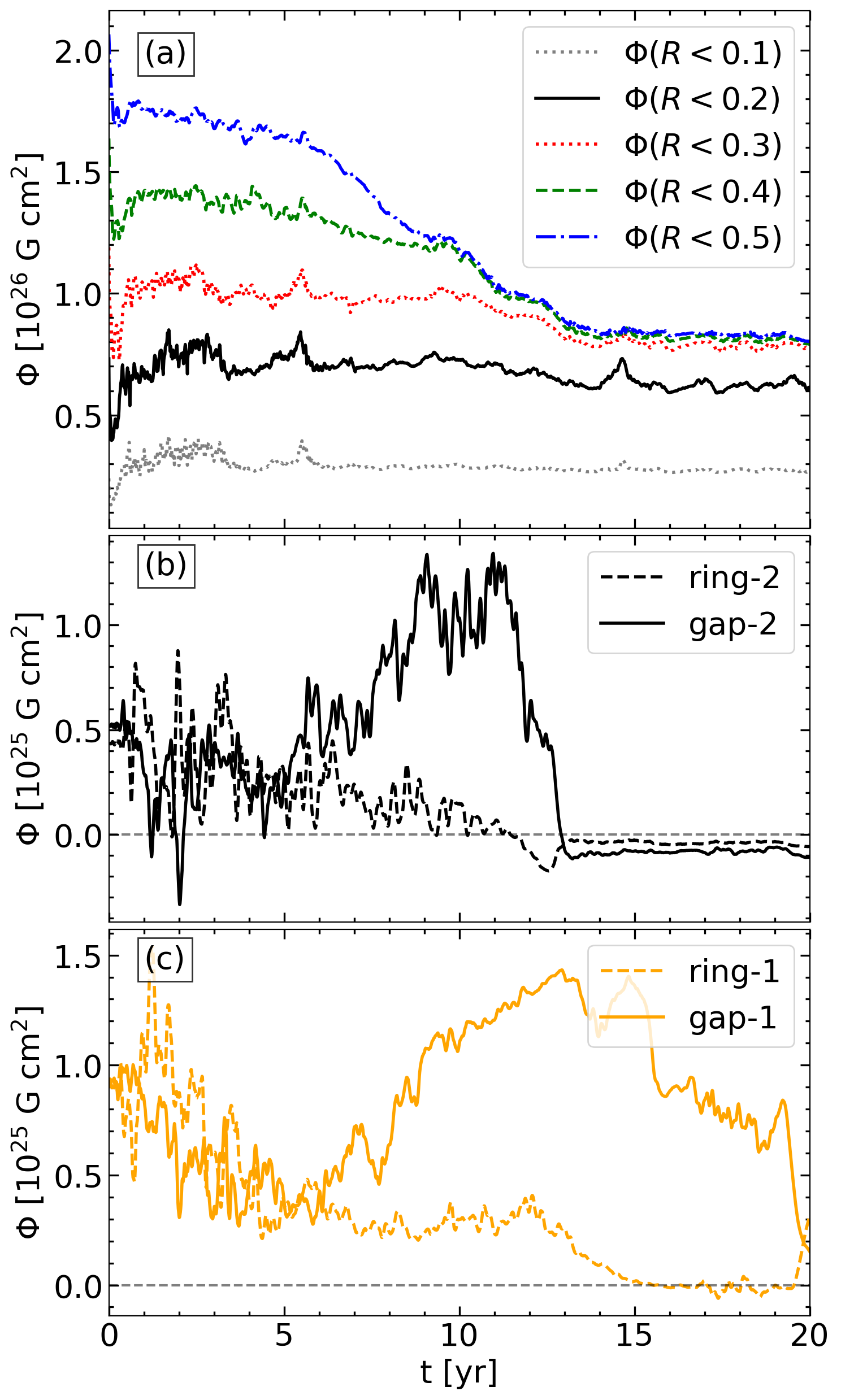}
    \caption{Magnetic flux evolution in the simulation. \textbf{Panel (a)} shows the cumulative flux (defined in equation~\ref{equ:mag_flux}) within 0.1, 0.2, 0.3, 0.4, and 0.5 au, respectively. \textbf{Panel (b)} shows the magnetic flux sampled within the Ring 2 (flux between $0.27~\mathrm{au}<R<0.28~\mathrm{au}$) and Gap 2 ($0.24~\mathrm{au}<R<0.25~\mathrm{au}$); \textbf{panel (c)} shows the flux sampled within Ring 1 ($0.20~\mathrm{au}<R<0.22~\mathrm{au}$) and Gap 1 ($0.16~\mathrm{au}<R<0.18~\mathrm{au}$), illustrating flux concentration in gap and the ``magnetic dam'' effect of the rings.}
    \label{fig:Bflux}
\end{figure}

Figure~\ref{fig:Bflux}(a) shows the evolution of the cumulative magnetic flux within different radii. By definition, the flux within 0.1 au contains only the contribution from the inner boundary, whereas the larger radii also include the magnetic flux threading the disk midplane. As expected, the total magnetic flux enclosed within 0.1 au and 0.2 au stays roughly constant in the simulation, as the disk in this region remains in the ideal MHD limit, preventing flux loss. In contrast, the total magnetic flux enclosed at larger radii decreases substantially during the simulation. 
The magnetic flux between 0.4--0.5 au, and later between 0.3--0.4 au, is gradually lost during the simulation, coinciding with the cooling of these regions and the consequent onset of strong non-ideal MHD effects
(see also Figure~\ref{fig:ring_gap}).

To further quantify the magnetic flux evolution and its connection to the ring-gap structures, Figures~\ref{fig:Bflux}(b) and (c) show the magnetic flux associated with Ring 2/Gap 2 and Ring 1/Gap 1, respectively (see Figure~\ref{fig:ring_gap} for definitions). 
Because the positions and widths of the rings and gaps evolve with time, we use narrow radial intervals to represent the magnetic flux within each structure. Specifically, the flux associated with Ring 2 is measured between $0.27~\mathrm{au}<R<0.28~\mathrm{au}$, while the flux in Gap 2 is measured between $0.24~\mathrm{au}<R<0.25~\mathrm{au}$. Similarly, Ring 1 and Gap 1 are represented by the radial intervals $0.20~\mathrm{au}<R<0.22~\mathrm{au}$ and $0.16~\mathrm{au}<R<0.18~\mathrm{au}$, respectively. In both ring-gap pairs, the magnetic flux preferentially concentrates within the gaps, whereas the rings contain comparatively little magnetic flux. This behavior is consistent with previous studies showing magnetic flux concentration in low-density gaps \citep[e.g.,][]{Zhu2013, Suriano2017, Riols2019, Hu2023, Hu2025}.

The degree of magnetic flux concentration in the gaps depends strongly on whether the adjacent exterior ring remains in the ideal-MHD regime. Around $t\sim13~\mathrm{yr}$, Ring 2 cools below the MRI activation temperature ($T_\mathrm{MRI}\approx800~\mathrm{K}$;\footnote{We note that MRI activation also depends on the gas density, magnetic field strength, and non-thermal ionization. Therefore, the adopted temperature threshold serves as a simplified diagnostic criterion rather than a strict condition for MRI activation.} Figure~\ref{fig:spacetime_and_accretion} and \ref{fig:heating_cooling_and_spacetime}). Once this occurs, magnetic flux is rapidly lost from Gap 2, leaving both the ring and gap with negligible residual magnetic flux.
(Small negative flux values arise from the loop-like magnetic field geometry induced by the presence of Ring 1; see the right panels of Figure~\ref{fig:ring_gap}).
A similar evolution occurs for Ring 1 and Gap 1: magnetic flux initially concentrates in Gap 1, but once Ring 1 begins to cool at $t\sim19~\mathrm{yr}$, the magnetic flux in the gap also begins to depart.

\begin{figure*}
    \centering
    \includegraphics[width=\linewidth]{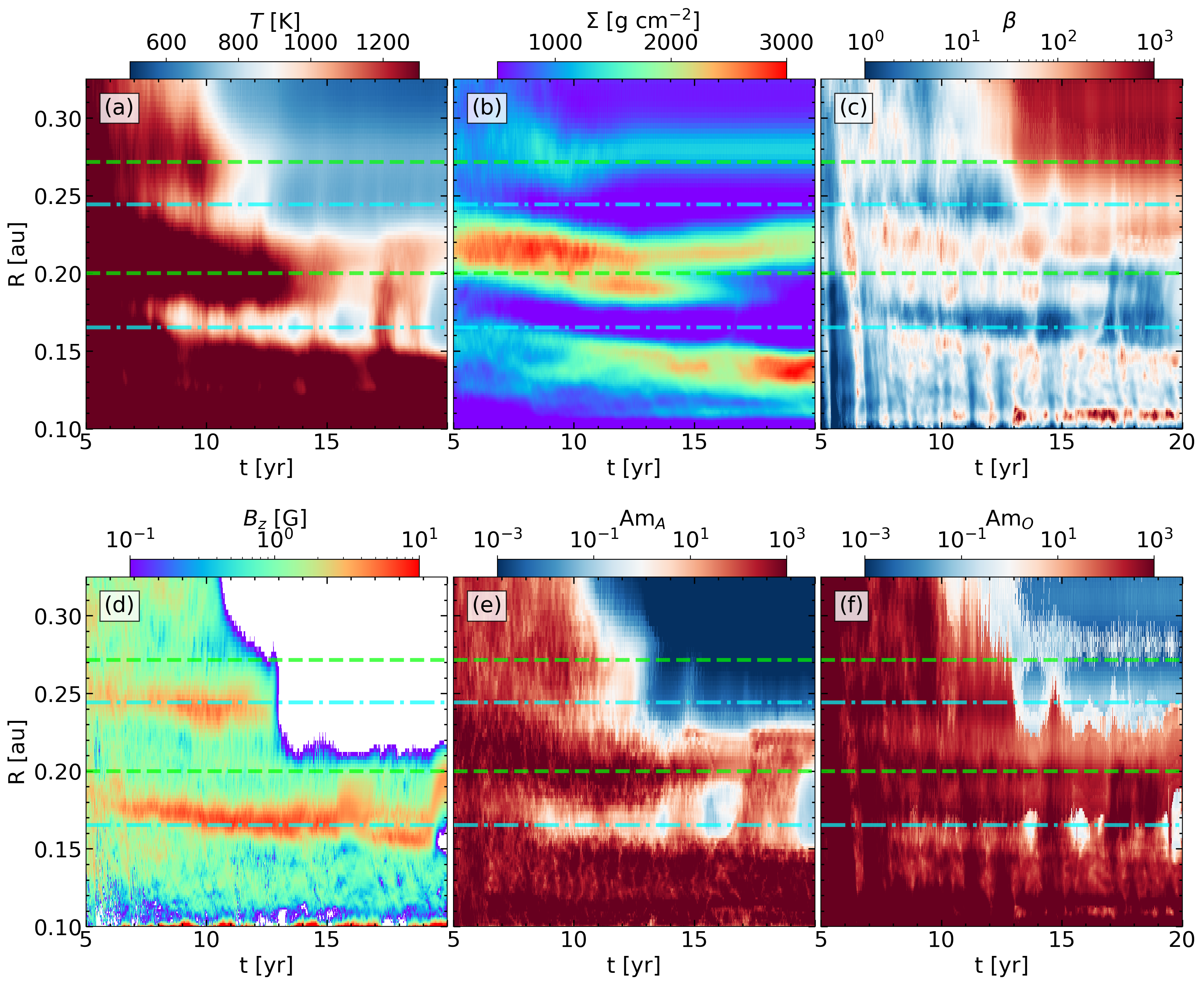}
    \caption{Time--space diagrams highlighting the ring decay and ``Reactivation'' events. \textbf{Panel (a)} shows the azimuthally and vertically averaged disk temperature, while \textbf{panel (b)} shows the azimuthally averaged disk column density. \textbf{Panels (c) and (d)} show the vertically averaged plasma-$\beta$ and vertical magnetic field strength $B_z$, respectively. The white regions in panel (d) correspond to locations where the averaged $B_z<0$ (see Figure~\ref{fig:ring_gap} for the magnetic field geometry and Section~\ref{sec:magfield_evol_ring_decay} for discussion). \textbf{Panels (e) and (f)} show the averaged Elsasser numbers associated with ambipolar diffusion and Ohmic dissipation, respectively. The horizontal green dashed lines mark the approximate locations of Ring 1 and Ring 2, while the cyan dash-dotted lines indicate the locations of Gap 1 and Gap 2.}
    \label{fig:SpaceTime_beta_Am}
\end{figure*}

To illustrate this interplay between gap flux concentration and flux loss following the cooling of the adjacent ring and present a more global picture, Figure~\ref{fig:SpaceTime_beta_Am} presents a zoomed-in, vertically and azimuthally averaged time--space diagram covering $t=8$--$17~\mathrm{yr}$, capturing the developement of Ring/Gap 1 and 2. Figures~\ref{fig:SpaceTime_beta_Am}(a) and (b) show the temperature and surface density, respectively. Two horizontal dashed green lines mark the approximate locations of Ring 1 and 2, while two horizontal dash-dotted cyan lines indicate the gaps 1 and 2 immediately interior to those rings. Ring 2 cools around $t\approx13~\mathrm{yr}$ when its temperature drops below $\sim800~\mathrm{K}$, the threshold required for MRI activity, despite remaining overdense. In contrast, Ring 1 bifurcates into two closely spaced rings around $t\approx12~\mathrm{yr}$ (Figure~\ref{fig:SpaceTime_beta_Am}[b]). The inner branch of this bifurcated structure gradually weakens in both surface density and temperature after $t\approx14~\mathrm{yr}$, eventually leaving only a single active ring.

The strong correlation between ring-gap locations and magnetic field concentration is illustrated clearly in Figures~\ref{fig:SpaceTime_beta_Am}(c) and (d), which show the plasma-$\beta$ and vertical magnetic field strength $B_z$, respectively. Along the gap locations (cyan dash-dotted lines), the plasma-$\beta$ is systematically lower than within the rings (green dashed lines), indicating stronger magnetic fields in the gaps. Correspondingly, $B_z$ is enhanced in the gaps. This magnetic flux concentration also produces larger effective $\alpha$ values within the gaps, as discussed in Section~\ref{sec:heating_and_cooling}.

To further characterize the flux loss process, Figures~\ref{fig:SpaceTime_beta_Am}(e) and (f) show the Elsasser numbers for ambipolar diffusion and Ohmic dissipation, respectively. Although the simulation is intrinsically multidimensional, we project the magnetic coupling into a one-dimensional framework for this analysis. Since the Elsasser number measures the ratio between the Lorentz force and the Coriolis force, we define the vertically averaged Elsasser number as
\begin{equation}
    \mathrm{Am}\equiv\frac{\int_\mathrm{disk}\frac{B^2}{4\pi\eta}dz}{\int_\mathrm{disk}\rho\Omega dz},
\end{equation}
where $B$, $\rho$, and $\Omega$ are the magnetic field strength, density, and angular velocity, respectively, and $\eta$ is the non-ideal MHD coefficient. We denote the Elsasser numbers associated with ambipolar diffusion and Ohmic dissipation by $\mathrm{Am}_A$ and $\mathrm{Am}_O$, respectively. Magnetic decoupling requires only one non-ideal MHD effect to dominate; once either Elsasser number falls below order unity, the magnetic field becomes effectively decoupled from the gas.

At the location of the outer ring, ambipolar diffusion becomes important at $t\approx12~\mathrm{yr}$, when $\mathrm{Am}_A\approx1$, coinciding with the onset of magnetic flux decay.\footnote{At first glance, this may appear contradictory: the gaps possess stronger vertical magnetic fields and lower surface densities, yet exhibit smaller Elsasser numbers. However, the total magnetic field strength is larger within the rings than within the gaps, owing to more efficient toroidal field generation inside the rings. The stronger total field strength and smaller non-ideal MHD coefficients therefore result in larger Elsasser numbers within the rings.} As the magnetic field weakens and MRI heating declines, ambipolar diffusion increasingly decouples the magnetic field from the gas, returning the region to a magnetically inactive state. A similar process occurs in the inner ring, where the inner portion becomes magnetically decoupled at $t\approx15.5~\mathrm{yr}$ ($\mathrm{Am}_A<1$), leading to a decrease in poloidal magnetic field strength as the flux spreads over a larger area, followed by a local reconcentration event (further discussion in section~\ref{sec:terminal_reactivation}). The remaining portion of the ring continues to act as a partial barrier to flux transport, although it is likely that this structure would also dissipate over longer timescales.

Overall, this behavior highlights the role of MRI-active rings as ``magnetic dams.'' 
While MRI-active, the rings maintain strong coupling between the gas and magnetic field, inhibiting radial flux transport and enabling magnetic flux to accumulate within the adjacent gaps. Once a ring becomes MRI-inactive, this barrier is removed, allowing the magnetic flux to diffuse outward.

\begin{figure*}
    \centering
    \includegraphics[width=\linewidth]{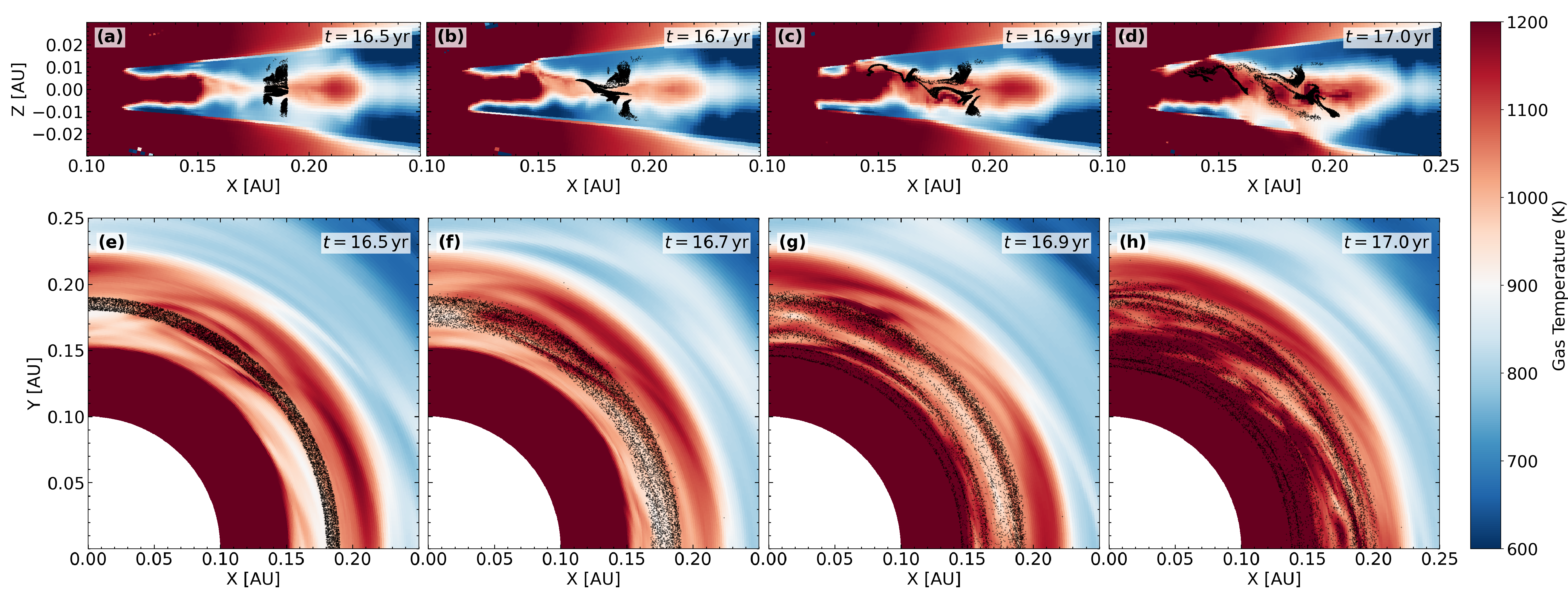}
    \caption{Temperature structure and gas dynamics during the ``Reactivation'' event. \textbf{The upper row (panels a--d)} shows the meridian-plane temperature evolution, with black dots marking post-processed tracer particles that trace the stream responsible for the reheating event. \textbf{The lower row (panels e--h)} shows the midplane temperature evolution, highlighting the passage of the stream and the resulting reheating.
}
    \label{fig:colweight_3D_temperature}
\end{figure*}

\subsection{Reactivation: local triggering events and outward ring migration}
\label{sec:terminal_reactivation}

One interesting feature during the decay phase is a brief reactivation of the inner disk around $t\approx17$ yr, which we refer to as a ``Reactivation'' event. During this event, part of the initially MRI-dead disk temporarily returns to the active state before cooling again.

To illustrate the evolution of this event, Figure~\ref{fig:colweight_3D_temperature} shows the gas dynamics traced by post-processed tracer particles. At $t=16.5$ yr, Ring 1 and Ring 2 are separated by a clear gap spanning $0.15~\mathrm{au}<R<0.18~\mathrm{au}$. Around $t=16.7$ yr (panels b and f), an accretion stream first develops near the disk midplane and then rapidly shifts toward one side of the disk. As this stream propagates through the gap (at $\sim16.9$ yr, panels c and g), it reheats the surrounding gas, and by $t=17$ yr (panels d and h) the inner disk has temporarily become MRI active again. After this brief reactivation, the heated region gradually cools and returns to the MRI-dead state toward the end of the simulation.

The physical origin of this ``Reactivation'' event is the partial decay of Ring 1 (Figure~\ref{fig:SpaceTime_beta_Am}[a]: the temperature of the inner part of Ring 1 decreases to $\lesssim T_\mathrm{MRI}$ at $t\sim15~\mathrm{yr}$). As the ring cools and becomes magnetically decoupled, the magnetic flux previously confined by the ring is released into the adjacent region (Figure~\ref{fig:SpaceTime_beta_Am}[d], where the $B_z$ concentrated in Gap 1 enters Ring 1 and increases the $B_z$ there around $\sim15.5~\mathrm{yr}$, preceding the ``Reactivation'' event around $\sim16.5~\mathrm{yr}$). The transient increase in magnetic flux temporarily reactivates MRI, producing the accretion stream and the accompanying reheating episode.

A secondary consequence of the ``Reactivation'' event is a modest outward migration of Ring 1 (Figure~\ref{fig:spacetime_and_accretion}[c]). Although the displacement is small in the reference 3D model, considerably larger migrations produced by the same mechanism are found in the longer-duration 2D simulations (Section~\ref{sec:2D}). The migration occurs because the temporary MRI reactivation re-establishes magnetic coupling between the gap and the neighboring ring, allowing magnetic stresses to transport angular momentum from the inner disk into Ring 1. This transient angular momentum transfer exerts a net outward force on the ring, causing it to migrate slightly outward before the system returns to the magnetically inactive state.

\section{Extrapolation with 2D models}
\label{sec:2D}
\subsection{Comparing 3D and 2D}
\label{sec:compare_2D_3D}
Before using the 2D models to extrapolate the results of the 3D simulations, it is necessary to quantify the differences between the two approaches and establish a baseline for detailed comparison. To this end, we performed a 2D simulation using the same physical parameters and initial conditions as the corresponding 3D model, enabling a direct comparison between the two cases.

\begin{figure*}
    \centering
    \includegraphics[width=\linewidth]{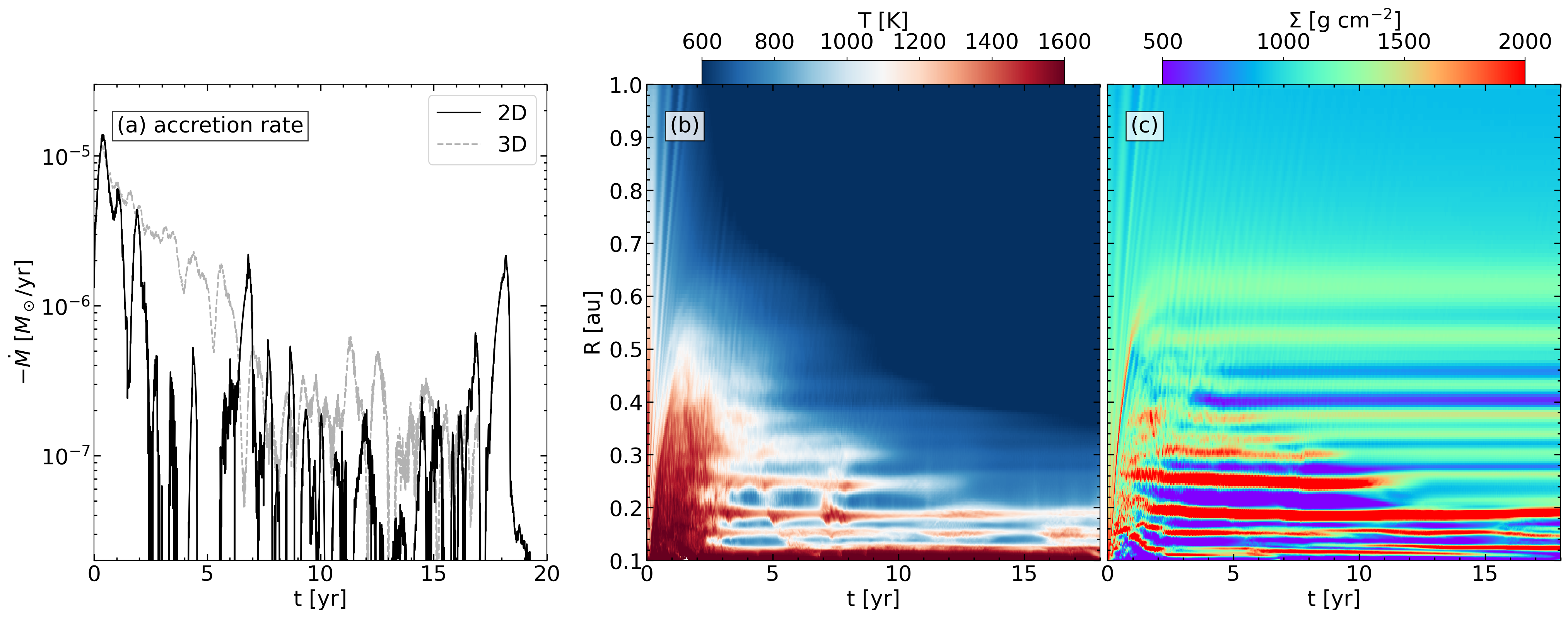}
    \caption{Accretion rate and time--space evolution of the 2D model, which is otherwise identical to the reference 3D model (Figure~\ref{fig:spacetime_and_accretion}). \textbf{Panel (a)} compares the accretion rates of the 2D and 3D models, with the 3D result shown as the gray dashed line. \textbf{Panels (b) and (c)} present the corresponding time--space diagrams of the disk temperature and column density for the 2D model.}
    \label{fig:2D_timespace_acc}
\end{figure*}

Figure~\ref{fig:2D_timespace_acc} presents the accretion rate, together with time-space diagrams of the column density and disk-averaged temperature. For reference, the accretion rate from the 3D model is shown as the gray curve in Figure~\ref{fig:2D_timespace_acc}[a]. Three major differences are apparent between the 2D and 3D simulations: (1) the 2D model develops more numerous rings and gaps than the 3D model; (2) the 2D model exhibits larger temporal variability; and (3) the outburst decays more rapidly in 2D. While differences (1) and (2) are expected and are well-known characteristics of 2D simulations \citep[e.g.,][]{Suriano2017, Suriano2019, Riols2020}, the third difference has not been recognized previously. 
In particular, in the 2D model, the accretion rate declines to a low state within $\sim4~\mathrm{yr}$, by which time the disk has become almost entirely magnetically decoupled beyond $\sim0.3~\mathrm{au}$. In contrast, the corresponding evolution in the 3D simulation requires approximately $\sim8~\mathrm{yr}$ to reach a similarly low accretion rate and comparable thermal structure.

\begin{figure*}
    \centering
    \includegraphics[width=\linewidth]{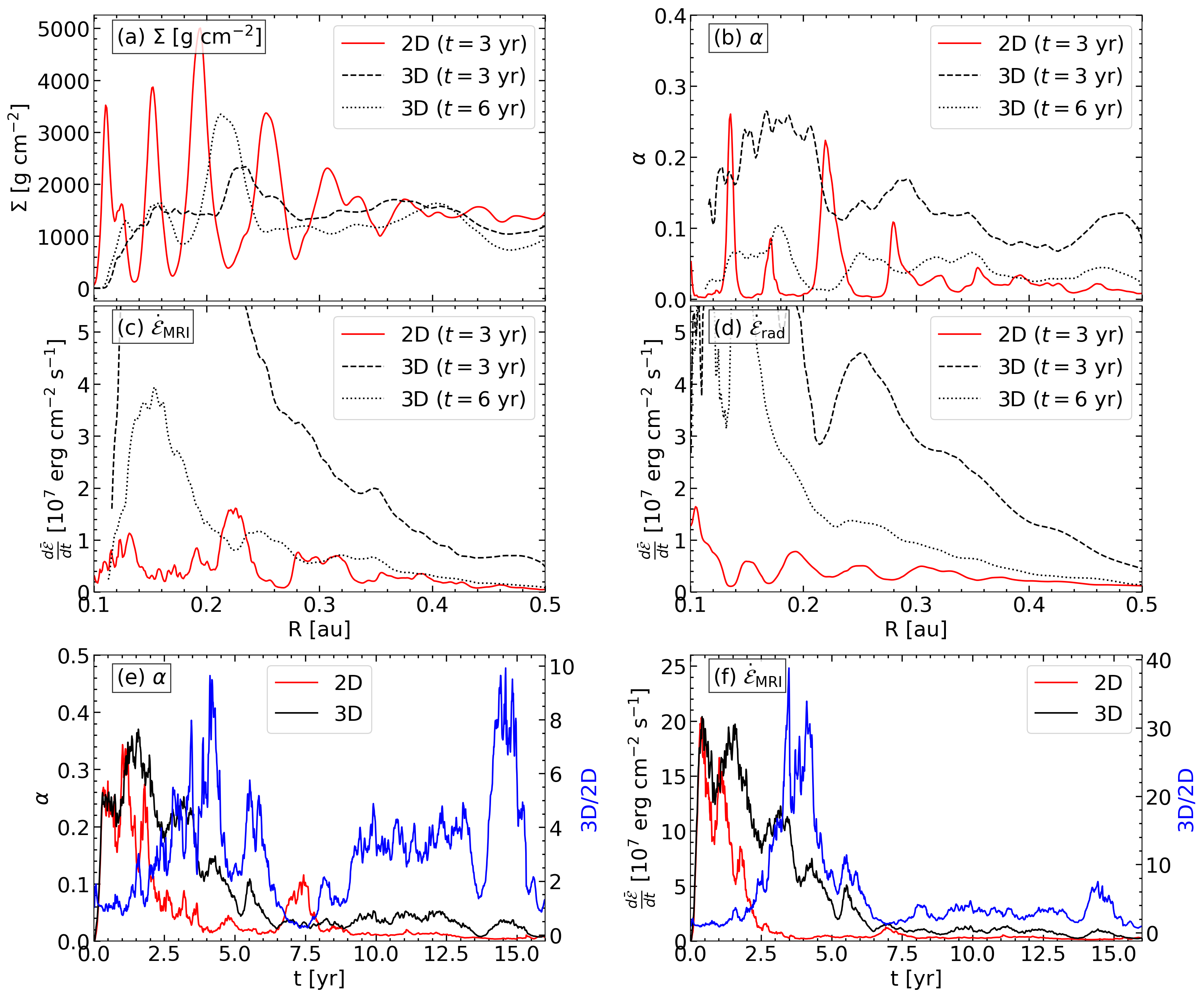}
    \caption{Quantitative comparison of the heating and cooling between the 2D and 3D models. \textbf{Panels (a) and (b)} show the disk column density and the effective viscosity parameter, $\alpha$ (Equation~\ref{equ:alphas}), respectively. 
    \textbf{Panels (c) and (d)} show the corresponding MRI heating and radiative cooling rates, respectively, computed using Equations~\ref{equ:dedt_MRI} and \ref{equ:dedt_rad}. \textbf{Panels (e) and (f)} show the time evolution of the azimuthally averaged $\alpha$ and MRI heating rate, respectively, averaged over $0.12<R<0.25~\mathrm{au}$, the region that remains hot throughout most of both simulations. The right-hand axes in panels (e) and (f) show the ratio between the 2D and 3D models.}
    \label{fig:heatcool2D3D}
\end{figure*}

To illustrate the more rapid decay in the 2D model compared to the 3D model, Figure~\ref{fig:heatcool2D3D} presents a quantitative comparison between the two at $t=3~\mathrm{yr}$, when the MRI has saturated and the 2D model is undergoing active decay. Because the temporal evolution of the 2D and 3D simulations is intrinsically different, constructing a strictly equivalent comparison between the two models is challenging. We therefore illustrate the difference by comparing the 2D and 3D simulations both at the same physical time ($t=3~\mathrm{yr}$) and against the 3D model at a later time ($t=6~\mathrm{yr}$), thereby partially accounting for the more rapid decay observed in the 2D case.

Since the thermal evolution determines how long the system remains in outburst, we examine the quantitative behavior of both MRI heating (equation~\ref{equ:dedt_MRI}) and radiative cooling (equation~\ref{equ:dedt_rad}). Figure~\ref{fig:heatcool2D3D}[a] shows the column density distribution in both models. 
The 2D simulation develops more numerous rings and gaps, with stronger density contrasts between them than in the 3D simulation. Correspondingly, the effective $\alpha$ parameter (Figure~\ref{fig:heatcool2D3D}[b]) exhibits significantly larger spatial variations in the 2D model. Importantly, however, aside from localized peaks near the gaps, the effective $\alpha$ in the 2D simulation is generally lower than in the 3D case, particularly within the rings where most of the mass resides. This difference is most pronounced in the same-time comparison, but remains evident even when compared to the later-stage 3D model.
As a consequence, the MRI heating rate in the 2D simulation remains systematically lower than in the 3D simulation.
This is quantified in 
Figures~\ref{fig:heatcool2D3D}[e] and [f] that show the area-weighted average $\alpha$ and heating rate $\frac{d\mathcal{E}}{dt}$ over the radial range $R=0.12$--$0.25~\mathrm{au}$. 
Overall, the effective $\alpha$ in the 3D
model exceeds that in 2D by roughly a factor of 3–4,
with the heating rate differing by a comparable amount.
This indicates that MRI heating operates substantially
more efficiently in 3D than in 2D. The lower temperature due to the lack of heating also reduces the effective radiative heating at the disk surface.
Nevertheless, with weaker heating, the cooling of the disk is faster, and the outburst decays faster.
In the following section (particularly section~\ref{sec:decay_time_scale}) we use the factor of $\sim 3$ decrease in decay time to
to extrapolate the long-term evolution inferred
from the 2D model to 3D.

\subsection{Overview of 2D models: long duration outbursts}
With the comparison between the 2D and 3D models and the limitations of the 2D models acknowledged in section~\ref{sec:compare_2D_3D}, we present a suite of three 2D models as a first step toward exploring the parameter space. Since our reference 3D model decays to a near-quiescent state after approximately 20 years, it is natural to ask what physical conditions are required to produce the much longer decay times observed in some FUor systems \citep[e.g.,][]{Fischer2023}. As established in Section~\ref{sec:heating_and_cooling}, the primary driver of outburst decay is radiative cooling.
It is therefore reasonable to expect that either increasing the disk density, thereby enlarging the thermal energy reservoir, or increasing the magnetic field strength, thereby enhancing MRI heating, could prolong the duration of an outburst.
We therefore explore these possibilities with a sequence of 2D models.

\begin{figure*}
    \centering
    \includegraphics[width=\linewidth]{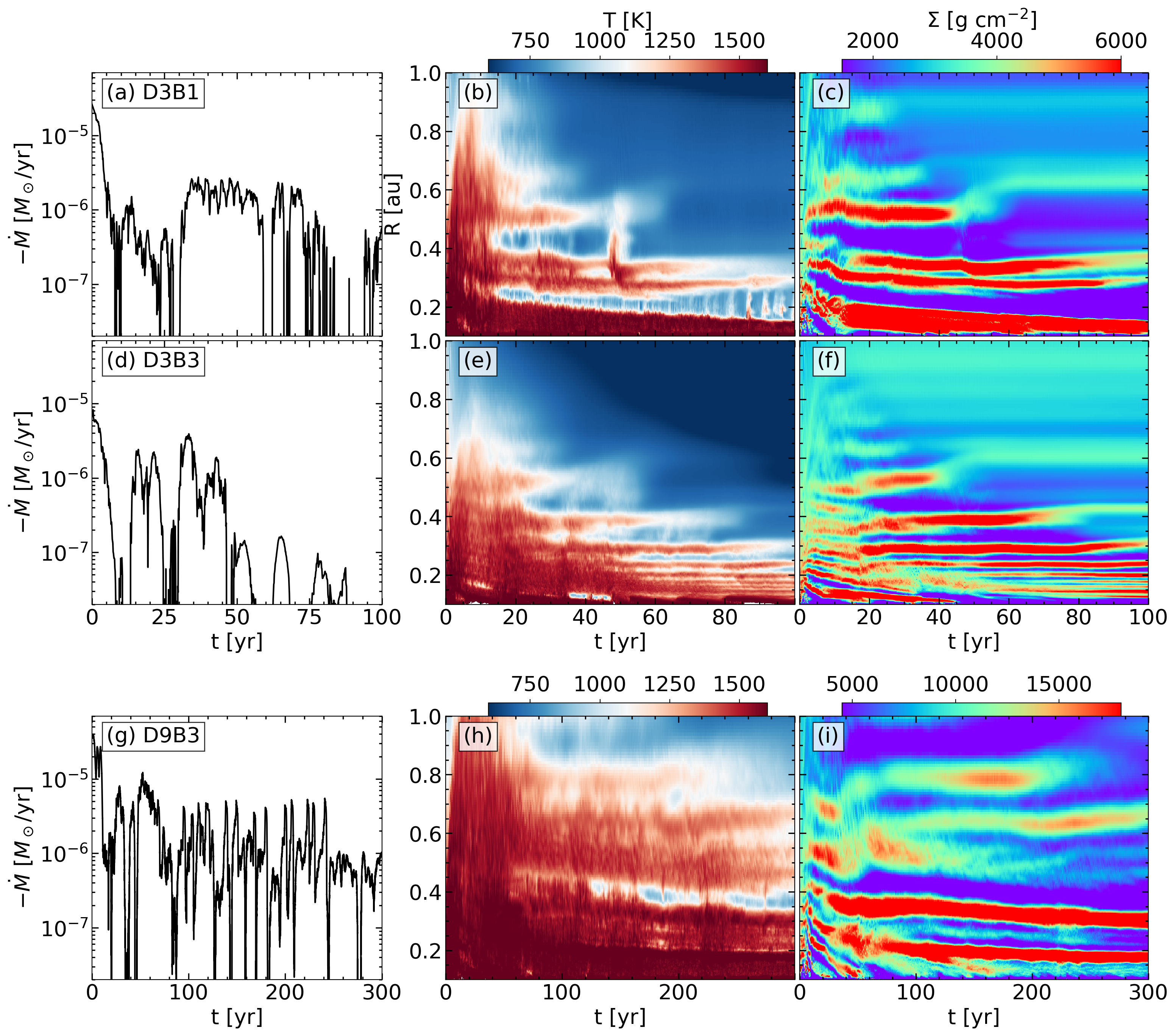}
    \caption{Overview of the long-duration 2D model suite. \textbf{The first row} shows the D3B1 model, in which the disk density is increased by a factor of three relative to the reference 3D model (Section~\ref{sec:3D_results}), while maintaining an initial midplane plasma-$\beta=1000$. The three panels show the accretion rate and the time--space diagrams of the disk temperature and column density, respectively. \textbf{The second row} shows the D3B3 model, which has the same disk density as the D3B1 model but an initial midplane plasma-$\beta=3000$, corresponding to the same magnetic field strength as the reference 3D model. \textbf{The third row} shows the D9B3 model, with a disk density nine times that of the reference 3D model and an initial midplane plasma-$\beta=3000$.}
    \label{fig:TimeSpace2DModels}
\end{figure*}

Figures~\ref{fig:TimeSpace2DModels}(a--c) show the results of the D3B1 model and Figures~\ref{fig:TimeSpace2DModels}(d--f) show the results of the D3B3 model, both of which have a density three times higher than the reference model and are evolved for 100 years, substantially longer than the 20-year evolution of the reference 3D model (Section~\ref{sec:3D_results}).

Interestingly, in both models, the MRI-active region remains confined within $\lesssim0.3~\mathrm{au}$ toward the end of the simulations. In the D3B1 model, the outermost active zone extends slightly farther outward than in the D3B3 model, although it is separated from the innermost active ring by a prominent gap near 0.2 au. In addition, the innermost ring in the D3B1 model continues to migrate inward throughout the simulation and eventually approaches the inner boundary. This inward migration helps sustain a relatively high accretion rate of $\sim2\times10^{-6}~M_\odot~\mathrm{yr}^{-1}$ over much of the simulation. Such behavior is expected, as the stronger magnetic field in this model facilitates angular momentum transport and ring migration. In contrast, inward migration is much weaker in the D3B3 model, occurring primarily during the early stages of the evolution ($\lesssim40~\mathrm{yr}$). Nevertheless, despite their different magnetic properties, both models retain MRI-active regions after a century of evolution.

The broadly similar long-term evolution of the D3B1 and D3B3 models suggests that an increased disk surface density plays the dominant role in sustaining long-duration outbursts, while the effect of initial vertical magnetic flux appears comparatively secondary. Indeed, both models remain active for far longer than the reference 3D simulation. To further illustrate the impact of disk density, we performed an additional model, D9B3 (Figures~\ref{fig:TimeSpace2DModels}[g-i]), in which the density is increased by a factor of nine relative to the reference model while maintaining an initial midplane plasma-$\beta=3000$, corresponding to the same vertical magnetic flux as in the D3B1 model. As expected, the MRI-active region in this model extends to significantly larger radii and remains active for substantially longer: after $\sim300~\mathrm{yr}$ evolution, only the outermost ring at 0.8~$\mathrm{au}$ decays, and the disk interior to 0.7~$\mathrm{au}$ is still hot and MRI active. The model also exhibits continued inward migration of active rings, indicating that the increased thermal reservoir associated with the higher surface density is highly effective at sustaining MRI activity and delaying the cooling of the outburst. Together, these results suggest that disk density is likely the primary factor controlling the duration of FUor outbursts, whereas variations in vertical magnetic flux mainly influence the detailed morphology and migration of the active rings.

\subsection{Thermodynamics in long duration 2D models}
\label{sec:2D thermo}
The decay of an outburst is governed by the balance between MRI heating and radiative cooling. In this section, we quantify this balance for the suite of long-duration 2D models.

\begin{figure*}
    \centering
    \includegraphics[width=\linewidth]{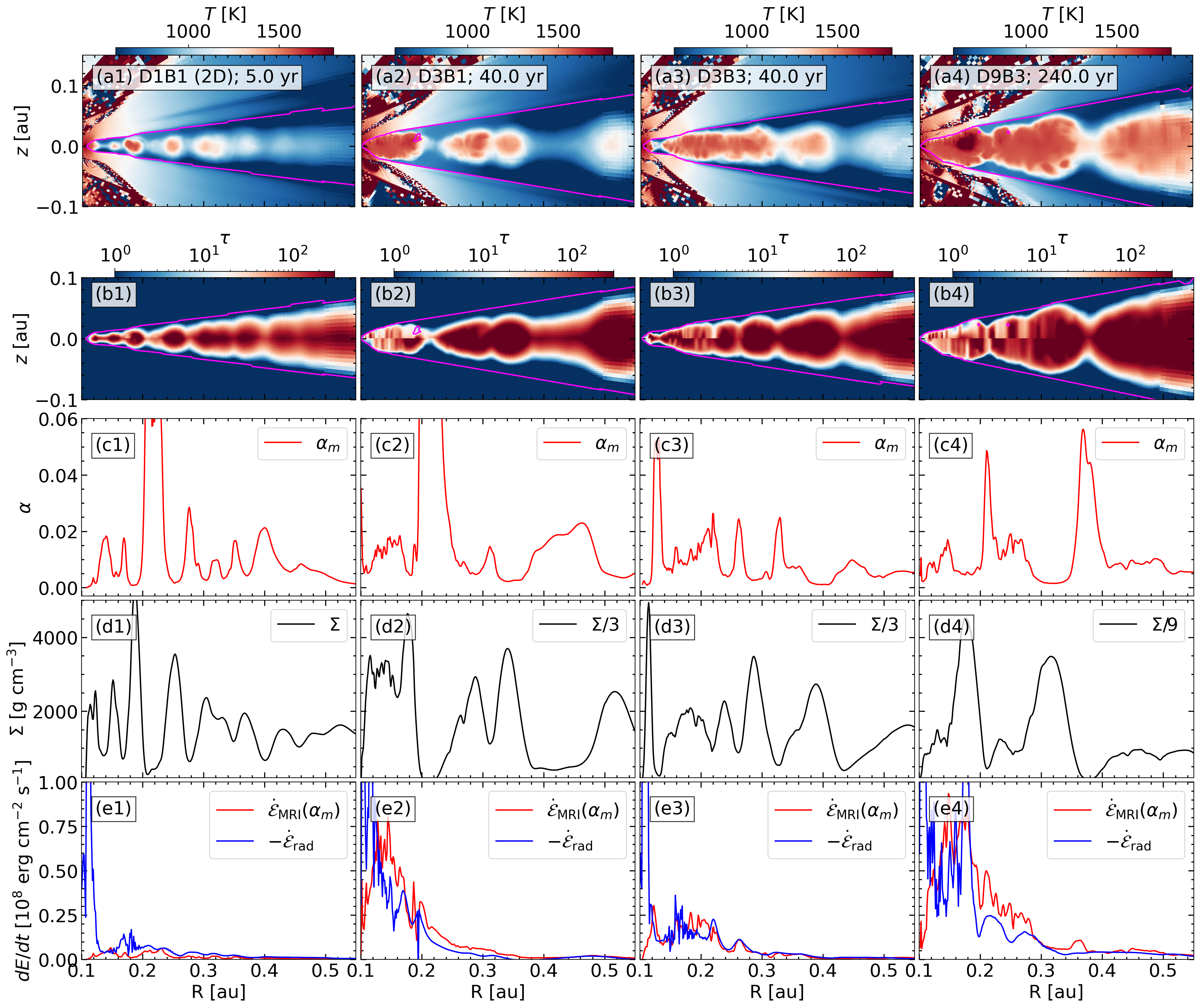}
    \caption{Meridional-plane structure and thermodynamic properties of the 2D models. \textbf{The first row (panels a1--a4)} shows the temperature distribution in the four 2D models. The purple contour outlines the disk, defined by $F_r<0.1~F_{r,~\star}$ (Section~\ref{sec:heating_and_cooling}). \textbf{The second row (panels b1--b4)} shows the disk optical depth, integrated vertically from the simulation boundary in each disk hemisphere. \textbf{The third row (panels c1--c4)} shows the vertically averaged effective viscosity parameter, $\alpha$ (Equation~\ref{equ:alphas}). \textbf{The fourth row (panels d1--d4)} shows the disk surface density, with the values in the D3B1, D3B3, and D9B3 models normalized to those of the D1B1 model for ease of comparison. \textbf{The fifth row (panels e1--e4)} shows the estimated MRI heating rate and radiative cooling rate, computed using Equations~\ref{equ:dedt_MRI} and \ref{equ:dedt_rad}, respectively.}
    \label{fig:2D_heat_cool}
\end{figure*}

We apply the same diagnostics introduced in Section~\ref{sec:heating_and_cooling}. Since the relevant quantities are vertically averaged, the formalism remains unchanged, although azimuthal averaging is no longer required in the 2D simulations. Figure~\ref{fig:2D_heat_cool} presents the vertical structure and thermodynamic properties of each model, with each column corresponding to a different simulation. Because the models evolve on different characteristic timescales, it is difficult to define a strictly synchronous comparison. Instead, we select a representative snapshot from each simulation during its cooling phase. While this choice is somewhat arbitrary, it enables a meaningful comparison of the thermodynamic state among the models.

Since radiative cooling depends sensitively on both temperature and optical depth, Figures~\ref{fig:2D_heat_cool}(a1--a4) and (b1--b4) show the temperature and opacity structures, respectively. The magenta line marks the boundary of the ``disk'' used in the vertical averaging procedure of Equations~\ref{equ:Tf} and \ref{equ:Pm}. 
At the selected times, all models exhibit well-developed ring-gap structures. The D1B1 model contains the smallest rings, while the D3B1 and D3B3 models show larger rings of comparable size. The D9B3 model exhibits the largest rings among the suite. A corresponding trend is seen in the temperature structure: ring temperatures increase from the D1B1 model to the D3B1 and D3B3 models, with the D9B3 model reaching the highest temperatures.

To characterize the opacity structure, Figures~\ref{fig:2D_heat_cool}(b1--b4) show the vertically integrated optical depth measured from the simulation boundary in each model. Interestingly, although the D3B1, D3B3, and D9B3 models have higher column densities than the D1B1 model, they exhibit lower optical depths, particularly in the inner disk. This is a direct consequence of the stronger MRI heating in these models, which raises the gas temperature above the dust sublimation threshold. The resulting reduction in dust opacity more than compensates for the increase in column density, producing a lower overall optical depth. This is analogous to the well-known ``thermostat effect'' of dust sublimation \citep[][]{DAlessio1998, Min2011, Woitke2024}.

To further quantify the enhanced MRI heating, Figures~\ref{fig:2D_heat_cool}(c1--c4) show the effective viscosity parameter, $\alpha$ (Equation~\ref{equ:alphas}). In all models, the rings maintain typical values of $\alpha\sim0.02$. Moreover, the trend identified in Section~\ref{sec:heating_and_cooling}—namely that $\alpha$ is systematically higher in the gaps than in the rings—persists throughout the entire model suite, providing further support for our interpretation that magnetic flux preferentially concentrates within the gaps.

A second crucial quantity controlling MRI heating is the disk surface density (Equation~\ref{equ:dedt_MRI}). To facilitate comparison among the different models, we normalize the surface density profiles of the higher-density models to that of the reference model. Interestingly, although the D3B1 and D3B3 models begin with identical density profiles, the rings in the D3B1 model develop systematically higher surface densities than those in the D3B3 model. In fact, the surface density distribution in the D3B1 model becomes qualitatively similar to that of the D9B3 model, which shares the same initial vertical magnetic flux. As discussed in Section~\ref{sec:magfield_evol_ring_decay}, magnetic flux preferentially concentrates in the gaps. A natural consequence of this flux concentration is that gas is expelled from the gaps and accumulates in the rings \citep[as reviewed by][]{Lesur2023}, further enhancing the ring surface density.

The combination of similar $\alpha$ values but different surface densities leads to substantial differences in the MRI heating rate, shown in Figures~\ref{fig:2D_heat_cool}(e1--e4) using Equation~\ref{equ:dedt_MRI}. As expected, the D1B1(2D) model exhibits the lowest heating rate. Comparing the D3B1 and D3B3 models, the D3B1 model shows significantly stronger MRI heating owing to its higher ring surface density, with heating rates comparable to those of the D9B3 model. It is therefore noteworthy that the D3B1 and D3B3 models exhibit similar outburst decay times despite their markedly different heating rates. The explanation lies in the radiative cooling.

As established earlier (Figures~\ref{fig:2D_heat_cool}[a--d]), the higher temperatures reached in the D3B1 and D9B3 models drive portions of the disk above the dust sublimation temperature. The resulting reduction in opacity makes the disk substantially more transparent to radiation, thereby enhancing radiative cooling. Thus, stronger MRI heating is accompanied by more efficient cooling.

\section{Discussion}
\label{sec:discussion}
\subsection{The long-lived aftermath of an FUor outburst}
\label{sec:long_live_aftermath}
Although an FUor outburst is observationally identified by its enhanced luminosity, the dynamical consequences of the event persist well after the system has returned to a quiescent appearance. In our simulations, the observational outburst effectively ends once the accretion rate and the extent of the hot MRI-active region have substantially declined (e.g., in the 3D model, at about 10 years, Figure~\ref{fig:spacetime_and_accretion}). However, this does not mark the end of the disk's evolution. Instead, the disk enters a prolonged relaxation phase during which the structures created during the outburst continue to evolve for decades.

One of the most important consequences of an outburst is the emergence of long-lived ring-gap structures. As discussed in Section~\ref{fig:ring_gap}, rings and gaps could form as a result of magnetic-field gas interaction, and the emergence of rings and gaps is therefore a natural part of the post-outburst evolution. Once formed, however, these rings do not disappear on the same timescale as the observable outburst. Although the innermost rings could be accreted (e.g., the innermost rings in the 2D models; Figure~\ref{fig:2D_timespace_acc}), many of the outer rings remain at nearly fixed radii throughout our simulations, even after their decay. The longevity of these rings arises in part because angular momentum is not removed solely through local accretion or disk winds. Instead, a significant fraction is redistributed through large-scale magnetic field lines connecting different radii within the disk, and such internal redistribution could even migrate some outer rings outward (e.g., section~\ref{sec:terminal_reactivation}). At the same time, magnetic flux becomes preferentially concentrated in the gaps rather than the dense rings themselves, reducing the magnetic braking within the rings. As the disk cools gradually, the rings transition back to the magnetically inactive state while largely retaining their mass. 

The longevity of these rings and gaps will depend upon the level of viscous diffusion \citep[][]{Lynden-Bell1974, Pringle1981}. Unfortunately, the value of an effective viscous $\alpha$ is extremely uncertain in non-magnetically active regions \citep[see][]{Lesur2023}. However, for a plausible dead zone $\alpha<10^{-3}$, the diffusion timescale can be upwards of $10^3$ years. Thus the
rings and gaps are likely to far outlast the outburst, and could be
important sites where the pressure maxima can concentrate solid particles and form planetesimals.

The disk evolution in our simulations differs qualitatively from previous
one-dimensional models of outbursts with or without thin active layers above otherwise dead zones
\citep[e.g.,][]{armitage01,Lodato2004,Zhu2010a,Zhu2010b, Nayakshin2024}. In 1D models, accretion resulting from a large increase in viscosity typically results in the accretion of most of the inner disk mass \citep[e.g.,][]{Bell1994,Lodato2004}. In contrast, our calculations show that
the accretion tends to happen along the disk surface.
The radiative trapping needed to keep temperatures above the activation level depends sensitively on the amount of mass above the accreting layers rather than upon the total surface density. 
In turn, this results in the MRI shutting off long before the surface density is strongly depleted.
Accretion thus only depletes part of the inner disk mass. This result
indicates that estimates of inner disk masses based on the total amount of mass accreted in an event \citep[e.g.,][]{Hartmann1996} may strongly underestimate true values. We also note that our setups and results differ physically from those of $\alpha$-disk models such as \citet[][]{Cecil2024, Cecil2026, Ziampras2026}, and that the ring-gap structures formed in their models differ fundamentally from those in ours. In their models, the rings form in association with reflares during the outburst, whereas our rings develop as a result of magnetic field concentration during a relatively smooth decline of the outburst. The rings in our models also persist for longer times, likely because the effective viscosity due to the magnetic field in our decayed dead zone is lower than that assumed in their $\alpha$-disk models. {Nevertheless, the rings formed by either mechanism can facilitate dust concentration and planetesimal formation.}

Moreover, our simulations suggest
that a substantial fraction of the disk mass remains confined to long-lived rings, which may be sites that promote planetesimal formation. As a result, the observable end of the outburst does not coincide with the end of the disk's dynamical evolution. The disk retains a structural memory of the outburst in the form of persistent rings and gaps that can survive long after the luminosity has returned to its quiescent level.

\subsection{What sets the decay time scale?}
\label{sec:decay_time_scale}
We have shown with a suite of 2D models that the outburst decay timescale varies substantially with disk surface density. It is therefore important to identify the key parameters that control the decay rate and, more generally, the duration of an outburst. While the answer inevitably depends on the initial conditions, particularly the extent of the disk activated during the triggering phase, we focus here on a more specific question: once a region has entered a self-sustaining MRI-active state, how long can it remain hot before returning to the magnetically inactive state?

In principle, the decay timescale can be defined as
\begin{equation}
    t_\mathrm{decay}\equiv\frac{E_\mathrm{disk}}
    {\left.\frac{dE}{dt}\right|_\mathrm{MRI}
    +\left.\frac{dE}{dt}\right|_\mathrm{rad}},
\end{equation}
where $E_\mathrm{disk}=\Sigma k_\mathrm{B}T/(\mu m_\mathrm{H})$ is the thermal energy of the disk, and the heating and cooling terms are given by Equations~\ref{equ:dedt_MRI} and \ref{equ:dedt_rad}, respectively. In practice, however, this expression is difficult to evaluate because MRI heating and radiative cooling remain close to equilibrium throughout most of the outburst evolution \citep[see also][]{Flock2013}. Furthermore, both terms depend nonlinearly on the disk temperature (Section~\ref{sec:2D thermo}). These complications caused us to seek a physically motivated, yet empirical, description of the decay timescale.



\begin{figure*}
    \centering
    \includegraphics[width=\linewidth]{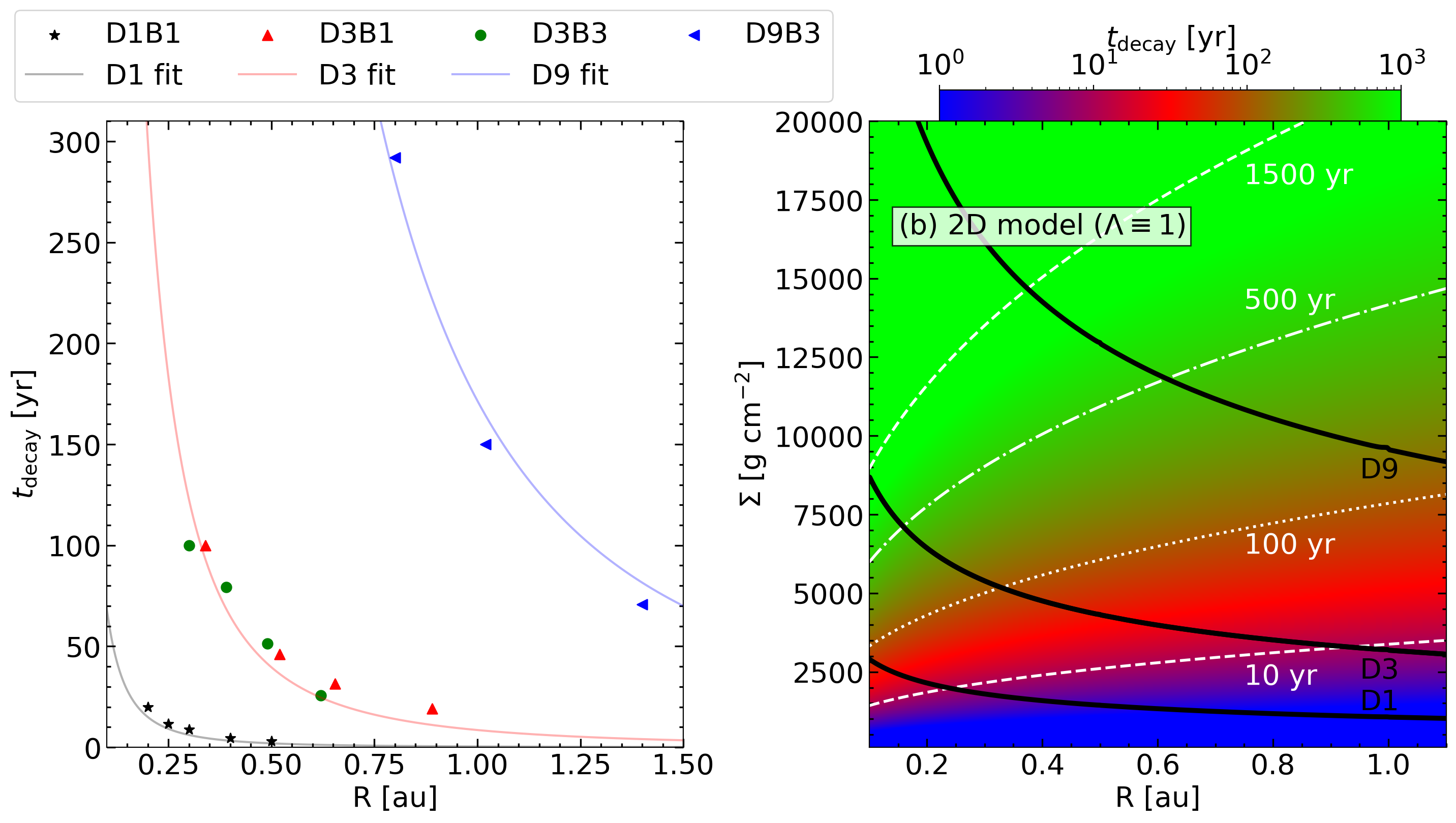}
    \caption{\textbf{The left panel} shows the estimated decay time at different radii for all four 2D models. Crosses indicate the values estimated by Equation~\ref{equ:tdecay}, demonstrating generally good agreement with the simulation results. \textbf{The right panel} shows the empirical relation given by Equation~\ref{equ:tdecay}, calibrated using the 2D simulations (with $\Lambda=1$).}
    \label{fig:tdecay}
\end{figure*}

Because ring-gap formation and the occasional ``Reactivation'' event introduce local variations, the decay time is not strictly a monotonic function of radius. We therefore estimate the local decay time as the epoch when each ring cools below $T\approx800~\mathrm{K}$, marking the loss of sustained MRI activity. The resulting decay times, measured at the ring locations, are shown in Figure~\ref{fig:tdecay}(a). The trends are clear: higher surface densities substantially increase the decay time, while the outer disk consistently decays earlier than the inner disk.

Interestingly, the magnetic flux and magnetic field strength themselves play only a secondary role once the disk has entered the MRI-active state. Although a sufficiently strong field is required to initiate and sustain MRI turbulence, the D3B1 and D3B3 models exhibit similar decay timescales despite their different initial magnetic field strengths. This suggests that, after MRI becomes self-sustaining, the outburst duration is governed primarily by the disk thermodynamics and surface density rather than by the magnetic field strength.

Motivated by these trends, we propose the following empirical power-law relation between the decay time, disk surface density, and radius,
\begin{equation}
    t_\mathrm{decay}
    =
    \Lambda~t_{\mathrm{decay},~0}
    \Big(\frac{R}{R_0}\Big)^u
    \Big(\frac{\Sigma}{\Sigma_0}\Big)^w,
    \label{equ:tdecay}
\end{equation}
where $\Sigma_0=6000~\mathrm{g~cm}^{-2}$ and $R_0=1~\mathrm{au}$ are the reference values, $t_{\mathrm{decay},0}=48~\mathrm{yr}$, $u=-1.02$, and $w=2.72$ are determined from the fit, and $\Lambda$ accounts for the fact that the relation is calibrated using 2D simulations whereas real disks are intrinsically three-dimensional (Section~\ref{sec:compare_2D_3D}). The fitted decay times are shown as crosses in Figure~\ref{fig:tdecay}(a), demonstrating good agreement with the simulation results (see also the stress tests in Appendix~\ref{app:stress}). Figure~\ref{equ:tdecay}(b) illustrates the corresponding dependence of the decay time on radius and surface density predicted by Equation~\ref{equ:tdecay}. Comparing the D1B1 (3D) and D1B1 (2D) models suggests $\Lambda\approx3$, although a more systematic exploration of parameter space with fully three-dimensional simulations will be required to calibrate this factor and this relation more robustly.

\subsection{Caveats and limitations}
\label{sec:limitation}
While our simulations represent an advance in the treatment of MRI evolution over prevous $\alpha$ disk models, there are limitations and omissions of our treatment. One possible limitation is the treatment of outflows. While investigations have indicated 
that magnetically-coupled winds can have important effects on angular momentum transport, driving accretion 
\citep[e.g.,][]{pudritz86,konigl89,bai13, Bai2016, Rodenkirch2020, Tu2025, Tu2026b, Mori2025}, the outflows in our simulations are weak, corresponding to $\lesssim1\%$ of the accretion rate.
Most of the ejected material is
launched along the back-swinging magnetic field lines (see Sections~\ref{sec:heating_and_cooling} and Figure~\ref{fig:ring_gap}) and eventually fall back onto the outer disk surface. 
Consequently, these returning outflows primarily redistribute mass and angular momentum within the disk rather than removing them from the system. 

The above-mentioned studies show that wind launching
depends upon the ionization of upper disk layers.
One process we have neglected is ionization by X-rays from the central star. In Appendix~\ref{sec:X-ray} we present a preliminary investigation of the influence of X-rays in a 2D model, finding only modest effects. More detailed investigation  
of these outflows and their role in redistributing disk material and angular momentum is deferred to future work that pushes the inner simulation boundary closer to the stellar surface.

Another issue is the neglect of the Hall effect.
Although Hall diffusion is not expected to play a dominant role in the MRI-active region of the disk, which remains close to the ideal MHD limit, it could introduce additional dynamics during the decay phase, when the disk has cooled but the magnetic field remains relatively strong \citep[e.g.,][]{Armitage2015}. The Hall effect is also sensitive to the polarity of the magnetic field relative to the disk rotation axis, introducing an additional degree of uncertainty \citep[e.g.,][]{Wardle1999, Bai2015}. Nevertheless, because the Hall effect is expected to become important only after the disk has cooled substantially, we do not expect it to qualitatively alter the overall evolutionary picture presented here. In particular, the formation of rings and gaps during the outburst, and the subsequent outward diffusion of magnetic flux once a ring cools below the MRI activation temperature, should remain robust, although it may modify the remaining level of turbulence and the timescale of decay.

Still another question involves the loss of magnetic flux.
Section~\ref{sec:magfield_evol_ring_decay} shows that the rings act as ``magnetic dams,'' concentrating magnetic flux in the gaps interior to them; when the ring decays, the magnetic flux in the gap floods out and is lost to the outer disk. One further implication of this finding is that the system is inherently susceptible to magnetic flux loss: the initial magnetic flux in our model is ``prescribed'' into the disk, and in test models initialized without an MRI-active zone, we find that magnetic flux diffuses outward efficiently, leading to a substantial reduction in the magnetic field strength throughout the inner disk. While such flux loss does not pose a problem for a single outburst event, it raises a more general question of how sufficient magnetic flux can be retained in the inner disk to sustain, and potentially trigger, repeated outbursts. This is a particular case of a more generally recognized problem for the evolution of magnetic flux in protoplanetary disks, the resolution of which is not clear 
\citep[see, e.g.,][]{bai26}, and may involve disk/star dynamo processes \citep[][]{Uzdensky2002,Matt2005,Romanova2009,Donati2009,Ferrario2015,Zanni2013,Zhu2025,Tu2026} and/or interactions with the surrounding environment or protostellar envelope \citep[][]{Vaisala2019,Machida2020,Tu2024}.

Finally, to properly test the simulations against observations of photometric and spectral evolution, as well as the thermal processing of inner disk material, it will be necessary to extend the computational domains to much smaller radii, as it is clear that during outburst the disk must extend inward near or to the stellar surface \citep{Hartmann1996}.  {The variations in accretion rate shown in Figures~\ref{fig:spacetime_and_accretion} and~\ref{fig:2D_timespace_acc}, and the localized sporadic ``Reactivation'' event (Section~\ref{sec:terminal_reactivation}) may translate into observable fluctuations or re-brightening in the light curve; however, the accretion luminosity is ultimately determined by the accretion onto the star, which cannot be captured without modeling the innermost disk and stellar surface.} Quantifying these will require additional simulations that include the interaction of the disk with the stellar magnetosphere.






\section{Conclusion} 
\label{sec:conclusion}
We present one three-dimensional radiation non-ideal MHD simulation of the decay of an FUor outburst, together with a suite of two-dimensional models that explore the parameter dependence of the decay. The main conclusions from the 3D simulation are as follows:
\begin{enumerate}
    \item Rings and gaps form naturally as a consequence of the decay of an FUor outburst.
    \item Magnetic flux preferentially concentrates in the gaps, leaving the rings comparatively weakly magnetized. As a result, the rings and gaps exhibit comparable MRI heating rates despite the substantially higher surface density of the rings.
    \item During the decay phase, the rings act as ``magnetic dams'' that temporarily confine magnetic flux within the adjacent inner gaps. When a ring cools below the MRI activation temperature, non-ideal MHD effects decouple the gas from the magnetic field, allowing the concentrated flux in the corresponding gap to rapidly diffuse outward. Consequently, both the ring and the gap return to a magnetically inactive state.
    \item During this rapid outward diffusion of magnetic flux, the transient increase in magnetic flux threading a neighboring ring can temporarily reactivate MRI, producing a localized reheating event that we term ``Reactivation''. In such an event, an almost magnetically dead ring briefly becomes MRI active again and undergoes a transient temperature increase before cooling once more and returning to a magnetically inactive state as the excess magnetic flux diffuses away.

\end{enumerate}

To place these results in a broader context, we further performed a suite of 2D models to compare with the 3D simulation and explore the parameter dependence of outburst decay. The main findings from the 2D model suite are as follows:
\begin{enumerate}
    \item By comparing an otherwise identical pair of 2D and 3D models, we find that MRI heating is less efficient in two dimensions, with the heating rate typically lower by a factor of a few. Empirically, this reduced heating shortens the outburst duration by a factor of approximately $3$--$4$.
    \item Long-duration simulations with higher disk surface densities demonstrate that the outburst decay time depends strongly on the disk surface density, with a doubling of the surface density increasing the outburst duration by roughly a factor of four. Motivated by this trend, we propose an empirical relation between the decay time and disk surface density (Equation~\ref{equ:tdecay}), while emphasizing that the relation is calibrated using 2D simulations and therefore includes a calibration factor to account for three-dimensional effects.
\end{enumerate}

Finally, future work will be needed to connect these models with the larger protostellar environment and to make more direct comparisons with observations.

\begin{acknowledgments}
We thank the anonymous referee for a constructive and detailed report. We thank Shangjia Zhang and Adolfo Carvalho for insightful discussions. YT was supported by NASA grant 80NSSC24K1285; LH and ZZ received partial support. ZZ acknowledges support from NSF award 2408207, 2429732. FJC acknowledges support by the National Aeronautics and Space Administration under agreement No. 80NSSC21K0593 for the program ``Alien Earths.'' Computing resources were provided by the 
EPL HPC at Carnegie Science Earth and Planets Laboratory Division; the
NASA High-End Computing (HEC) Program through the NASA Advanced Supercomputing (NAS) Division at Ames Research Center; Anvil Supercomputer at Purdue University, Stampede3 Supercomputer at the University of Texas, made available through the NSF computing allocation PHY-250048, PHY-260035; and the computational resources and services provided by Advanced Research Computing at the University of Michigan, Ann Arbor. 
\end{acknowledgments}

\software{Athena++\citep[][]{Jiang2014, Stone2020}, NumPy \citep{Numpy}, Matplotlib \citep{Matplotlib}, Numba \citep{numba}}


\appendix

\section{Numerical Resolution Study}
\label{sec:convergence}
The 3D simulations are computationally expensive, both in terms of CPU core-hours and wall-clock time. We therefore adopt the minimum resolution required to adequately resolve the MRI \citep{Hawley2011}, corresponding to approximately eight cells per disk scale height at $R\sim0.4~\mathrm{au}$, where the outburst decay is most rapid. Given this choice, it is important to assess the sensitivity of our results to numerical resolution and determine whether the main conclusions would change at higher resolution.

Since performing a higher-resolution 3D simulation is currently impractical, we instead use 2D models to examine numerical convergence. This approach is motivated by the fact that the primary resolution requirement arises from adequately resolving the MRI on the meridional plane. The D1B1(2D) model (Figure~\ref{fig:2D_timespace_acc}) uses the same meridional resolution as the reference 3D model and therefore provides a natural baseline for comparison. We perform an additional higher-resolution simulation with identical physical parameters but increase the disk refinement level from 3 to 4. This doubles the effective spatial resolution in the disk and yields approximately 16 cells per scale height at 0.4 au.

\begin{figure*}
    \centering
    \includegraphics[width=\linewidth]{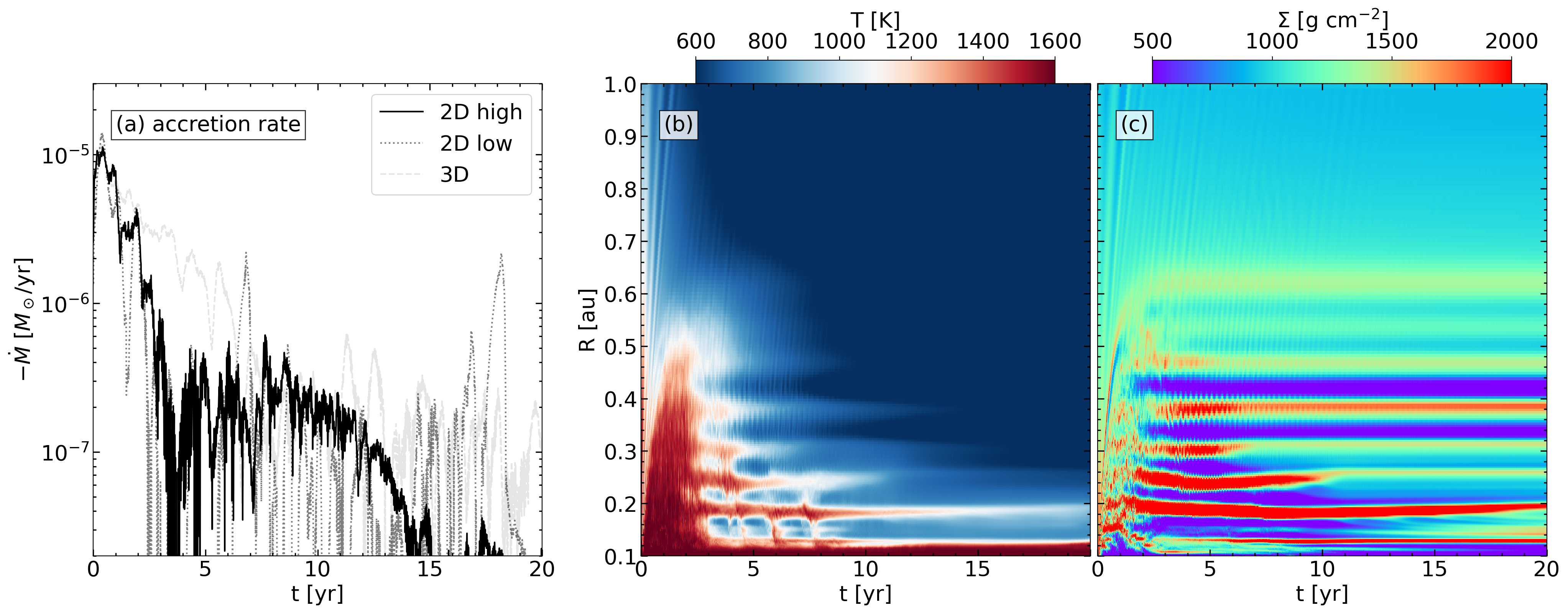}
    \caption{The same model as the D1B1 (2D) model, but at higher spatial resolution (to be compared with Figure~\ref{fig:2D_timespace_acc}). \textbf{Panel (a)} shows the accretion rate of the high-resolution model (solid black line), with the D1B1 (2D) model and the reference 3D model overplotted in the background for comparison. Relative to the D1B1 (2D) model, the high-resolution model exhibits smaller temporal variations in the accretion rate, while the overall decay occurs on a similar timescale. \textbf{Panels (b) and (c)} show the time--space diagrams of the disk temperature and column density, respectively. Compared with the D1B1 (2D) model, the high-resolution model develops slightly fewer rings. Nevertheless, the overall evolution and outburst decay remain similar between the two models, indicating that our main results and conclusions are robust against numerical resolution.}
    \label{fig:2D_resolution_study}
\end{figure*}

Figure~\ref{fig:2D_resolution_study}(a) compares the accretion histories of the low-resolution 2D model, the high-resolution 2D model, and the reference 3D model. The high-resolution 2D simulation exhibits an accretion history very similar to that of the low-resolution model, particularly during the rapid cooling phase ($t\lesssim5~\mathrm{yr}$). At later times ($t\sim10~\mathrm{yr}$), the high-resolution model shows somewhat reduced temporal variability in the accretion rate, although the overall evolutionary trend remains unchanged. The temperature and surface-density spacetime diagrams of the high-resolution model are shown in Figures~\ref{fig:2D_resolution_study}(b) and (c), respectively, and can be compared directly with Figures~\ref{fig:2D_timespace_acc}(b) and (c).

One noticeable difference is that the low-resolution model produces slightly more rings (approximately 10) than the high-resolution model (approximately 8). This behavior may reflect enhanced growth of zonal flows at lower resolution \citep[][]{johansen2009}. Nevertheless, both simulations develop multiple rings and gaps, and their global evolution remains remarkably similar. In both cases, the outer active disk cools rapidly, retreating to $R\lesssim0.3~\mathrm{au}$ within approximately 7 years, while the innermost $\sim0.2~\mathrm{au}$ remains MRI-active throughout the duration of the simulation. We therefore conclude that increasing the numerical resolution can modify some details of the ring-gap structure, but the overall thermodynamic evolution, outburst decay behavior, and principal conclusions of this work are robust against numerical resolution.

\section{Additional parameters: X-ray model}
\label{sec:X-ray}
One physical process that we have necessarily neglected is ionization by stellar X-rays. During an FUor outburst, however, the stellar X-ray luminosity is poorly constrained observationally \citep[][]{Skinner2006, Skinner2010}, and the X-ray flux reaching the disk beyond $\sim0.1~\mathrm{au}$ is itself uncertain. It likely depends sensitively on the disk structure interior to $0.1~\mathrm{au}$, which may extend all the way to the stellar surface and could be substantially elevated by magnetic support and/or thermal expansion. Nevertheless, because X-ray ionization can enhance magnetic coupling, it is important to assess its potential impact on the outburst decay.

To estimate this effect, we follow the prescription in \citet[][]{Bai2009}, where the X-ray ionization rate is approximated using the gas column density between each cell and the star (represented by the inner radial boundary in our simulations). As a proof of concept, we adopt an X-ray luminosity of $10^{29}~\mathrm{erg~s^{-1}}$ and repeat the D1B1 (2D) model with X-ray ionization included.

\begin{figure*}
    \centering
    \includegraphics[width=\linewidth]{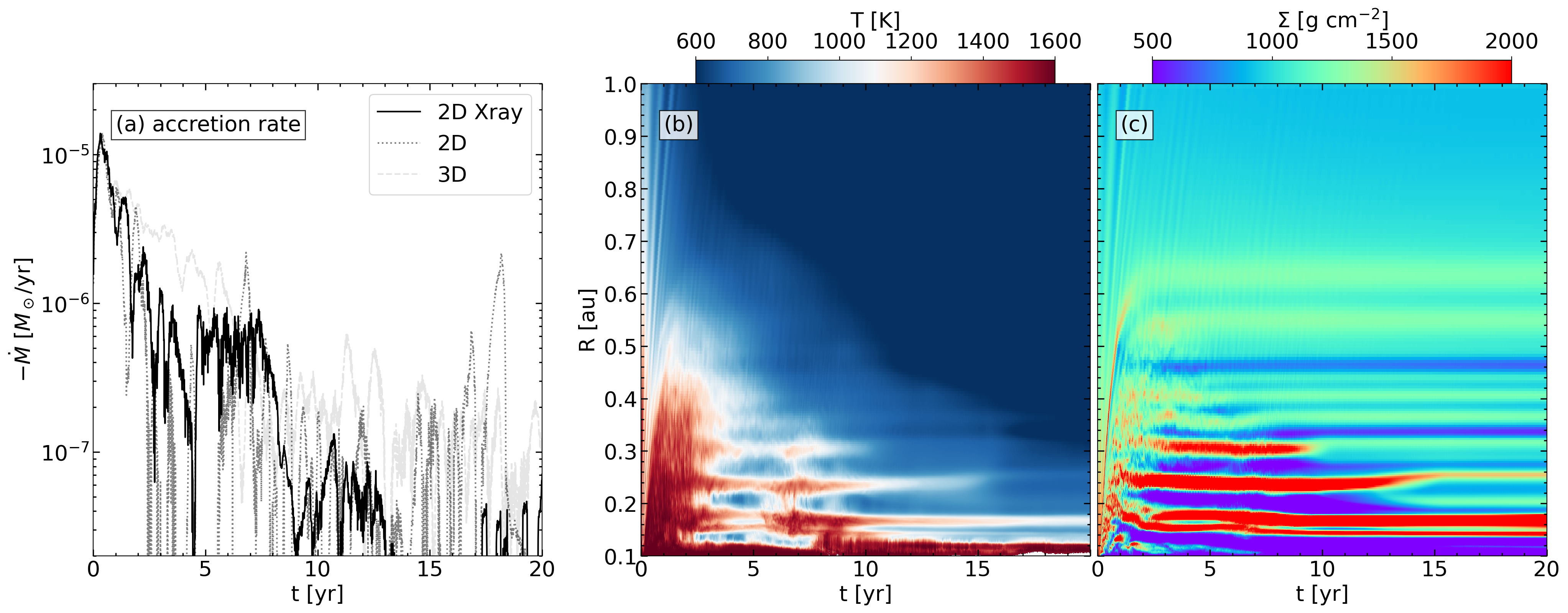}
    \caption{The same model as the D1B1 (2D) model, but with prescribed X-ray ionization (to be compared with Figure~\ref{fig:2D_timespace_acc}). \textbf{Panel (a), (b) and (c)} show the accretion rate and time--space diagrams of the disk temperature and column density, respectively.}
    \label{fig:2D_Xray_study}
\end{figure*}

Figure~\ref{fig:2D_Xray_study} compares the accretion rate and the time--space diagrams of the disk temperature and column density for the X-ray model. Overall, the evolution is remarkably similar to that of the reference D1B1 (2D) simulation. The only noticeable difference is a modest enhancement of the accretion rate between $t\approx5$ and $10~\mathrm{yr}$. This behavior is expected, as the additional ionization improves magnetic coupling in the inner disk, allowing the innermost rings to migrate inward more efficiently and thereby modestly enhancing accretion. However, the peak accretion rate, the decay timescale, and the overall evolution of the outburst remain essentially unchanged. We therefore conclude that stellar X-ray ionization has only a limited influence on the decay of an FUor outburst.

\section{Stress test with models with different density profiles}
\label{app:stress}
\begin{figure*}
    \centering
    \includegraphics[width=\linewidth]{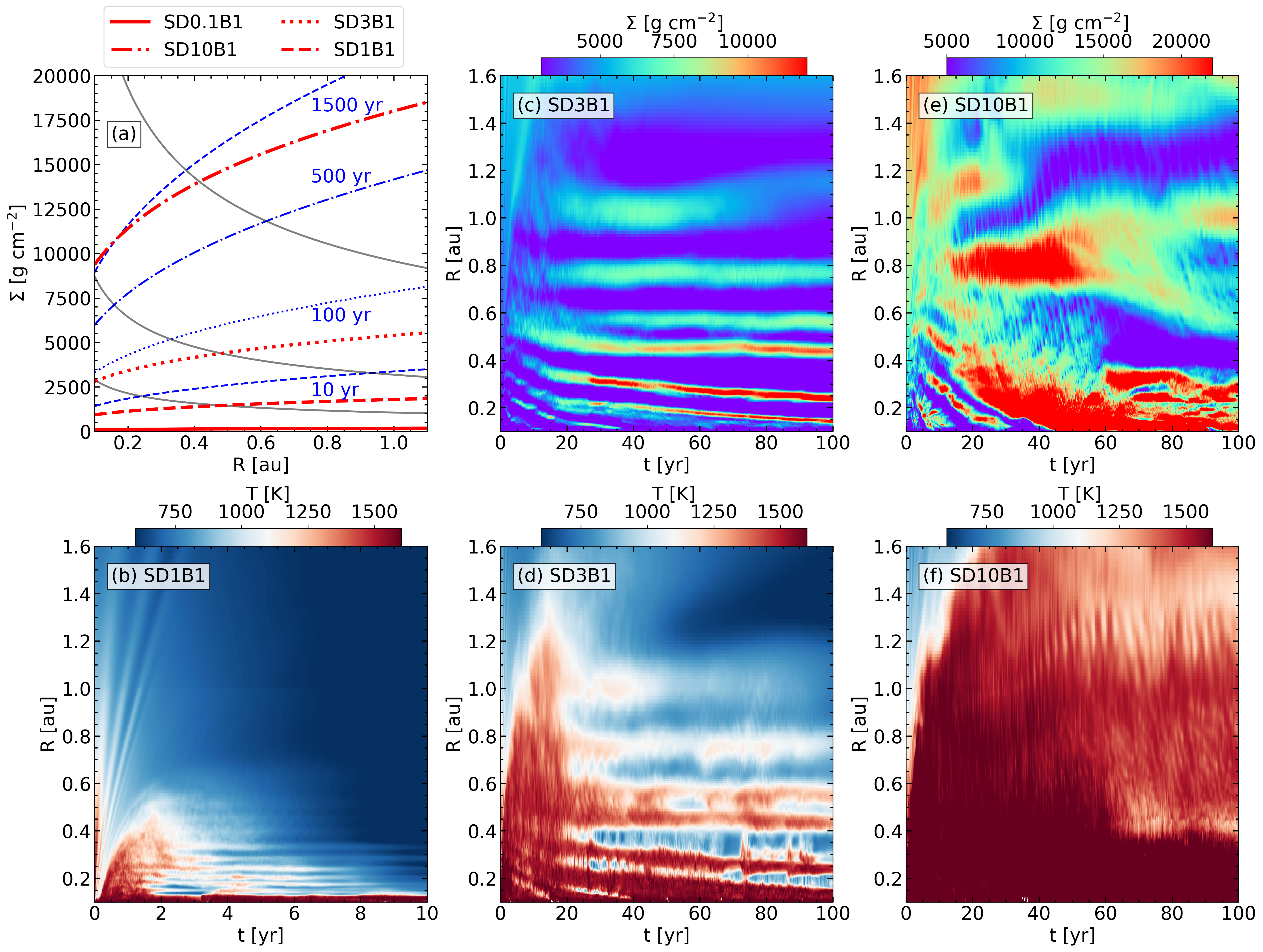}
    \caption{The Stress Test models, which have different density profiles from the main model suite. {\bf Panel (a)} shows the column density in each model, along with the $t_\mathrm{decay}$ estimation contours from equation~\ref{equ:tdecay} and section~\ref{sec:decay_time_scale}, figure~\ref{fig:tdecay}. {\bf Panel (b)} shows the temperature evolution in the SD1B1 model, showing a rapid decay. {\bf Panels (c) and (d)} show the column density and temperature evolution of the SD3B1 models, and {\bf panels (e) and (f)} show the evolution of the SD10B1 models, respectively.}
    \label{fig:stress_test}
\end{figure*}

Since one of the main findings of our model suite is the dependence of the decay timescale on disk surface density, it is important to test whether this result is robust to variations in the initial density profile. Here, we present four additional models as stress tests of the adopted density profile. These models are otherwise identical to the 2D models discussed in Section~\ref{sec:2D}, except that the initial density profile (Equation~\ref{equ:init_rho_T}) is modified to $q=-0.98$, resulting in a column density that increases slightly with radius. Although this surface-density profile is not particularly realistic, it provides a useful stress test of the robustness of our numerical results. We refer to these models as SD0.1B1, SD1B1, SD3B1, and SD10B1, corresponding to initial densities of 0.1, 1, 3, and 10 times the reference density, respectively. The initial column density profiles of these models are shown in Figure~\ref{fig:stress_test}(a), together with the same decay-timescale contours shown in Figure~\ref{fig:tdecay}(b).

The SD0.1B1 model provides a stress test of the low-surface-density regime, with a substantially lower density than the other models. In this model, there is effectively no distinct ``outburst'' stage, as the entire disk cools rapidly following initialization. We therefore focus on the remaining three models.

The SD1B1 model has a surface density similar to that of the D1B1 model in our main model suite, with a somewhat lower column density at smaller radii and a slightly higher column density at larger radii. Importantly, its surface density remains below the $t_{\rm decay}=10$ yr contour. The simulation result is shown in Figure~\ref{fig:stress_test}(b), and is consistent with this estimate: the outburst decays on a timescale of only a few years.

The SD3B1 and SD10B1 models are shown in Figures~\ref{fig:stress_test}(c,d) and~\ref{fig:stress_test}(e,f), respectively. Both models exhibit substantially longer decay timescales, further supporting our conclusion that a higher surface density leads to a longer-lasting outburst. The SD3B1 model lies below the $t_{\rm decay}=100$ yr contour, and the disk has indeed cooled substantially by 100 yr. However, some hot rings remain at this time. These rings have enhanced surface densities relative to the initial disk and therefore are expected to have longer local decay timescales. Similarly, the SD10B1 model lies above the $t_{\rm decay}=500$ yr contour and shows no significant cooling by 100 yr, again consistent with the expected dependence on surface density. {We note that, despite their substantially longer durations, the temperatures in the stress-test models remain around or below approximately 1500 K due to the ``thermostat effect'' of dust sublimation (see Section~\ref{sec:2D thermo}). The only exception is the very innermost region of SD10B1, where the temperature may exceed this value because of the excessively high density, similar to the behavior seen in model D9B3 in the main model suite.}

These stress-test models further support our conclusion that the outburst decay timescale depends sensitively on the disk surface density and that our estimated decay timescales are robust to reasonable variations in the initial density profile. They also highlight an important consequence of ring--gap formation: the enhanced surface density within the rings can locally prolong the hot state and thereby sustain the outburst for substantially longer than would be expected from the initial surface density alone.

\bibliography{sample701}{}
\bibliographystyle{aasjournalv7}



\end{document}